# Specification-Driven DevOps for Multi-Service Environments

Oleg Grynets
EPAM Systems
McLean, Virginia, USA
oleg_grynets@epam.com

Kyrylo Fursov
EPAM Systems
Lviv, Ukraine
kyrylo_fursov@epam.com

Vasyl Lyashkevych
EPAM Systems
Lviv, Ukraine
vasyl_lyashkevych@epam.com

Volodymyr Veres
EPAM Systems
Lviv, Ukraine
volodymyr_veres@epam.com

*Abstract*—**Large Language Models (LLMs) are increasingly used to generate executable software environments from repository artifacts. However, functional executability does not necessarily imply conformity with architectural, security, workflow, and production intent. This study investigates whether a frontier LLM can generate Dockerfiles and Docker Compose configurations for multi-service applications using repository contents without access to developer-authored deployment artifacts. Three heterogeneous repositories combining Python, Node.js, .NET, React, Rust, Java, Redis, PostgreSQL, and MySQL-compatible infrastructure were evaluated using deterministic end-to-end HTTP oracles and manual structural comparison. All three generated environments became functionally operational, although one required a Rust base-image update from version 1.85 to 1.88. The model correctly reconstructed service topology, application ports, infrastructure dependencies, service hostnames, a background worker, hidden proxy configurations, and a file-based Docker secrets mechanism. However, it consistently omitted network segmentation, multi-stage builds, dependency-layer caching, live-reload volumes, production frontend serving, restrictive backend-port policies, and cross-platform build logic. Based on these observations, the study formalizes the distinction between functional correctness and deployment-intent fidelity and analytically derives a minimal explicit deployment specification for information that cannot be reliably inferred from repository artifacts.**

## I. Introduction

The growing use of LLMs in software engineering is transforming the role of specifications within the software development lifecycle. LLMs are currently being applied to program synthesis, code autocompletion, testing, defect repair, requirements analysis, documentation generation, architectural reconstruction, and software migration [1]–[8]. However, most existing approaches still treat specifications as auxiliary natural-language prompts rather than as persistent, machine-processable engineering artifacts that govern the transformations between requirements, implementation, testing, deployment, and operation. Consequently, generated artifacts may be syntactically correct and even executable, yet remain inconsistent with implicit architectural, security, operational, or organizational intent.

This limitation is particularly significant in DevOps. Deployment environments combine application-specific knowledge with infrastructure decisions regarding service topology, runtime dependencies, network isolation, host access, secrets, image versions, platform constraints, build stages, observability, resource limits, and production operation and maintenance. Some of this information can be found in application repositories: package manifests indicate dependencies, source code reveals ports and hostnames, configuration files show database connection parameters, and proxy declarations point to inter-service interactions. However, other aspects—such as whether a service should be externally accessible, whether a network barrier is required, or whether the target environment is intended for development or production—are not necessarily reflected directly in the implementation.

Existing research on automated environment creation has demonstrated that repository content can support dependency inference and Dockerfile generation. Previous approaches identified operating system and language dependencies using static analysis, project structure, package metadata, and extracted configuration patterns [9]–[12]. Newer LLM-based methods utilize iterative feedback from build and test processes to create executable environments for individual repositories [13]–[17]. Repo2Run, for instance, iteratively builds Docker images and runs repository tests, achieving an 86% success rate across 420 Python repositories [13]. Subsequent developments have extended this approach to support various programming languages, the creation of multi-agent environments, and the use of dynamic repositories and advanced execution oracles [14]–[17]. However, these studies primarily focus on single-container environments capable of running an existing test suite. Typically, they do not address complete multi-service deployment environments, inter-service communication, network segmentation, credential passing, access policies for public ports, or the distinction between an executable environment and a deployment configuration that reflects the developer's intent [18].

A related line of research focuses on the generation of Infrastructure as Code (IaC) artifacts using LLMs. Testing results on datasets such as IaC-Eval, IaCGen, Multi-IaC-Eval, and others indicate significant improvements in syntactic correctness, whereas ensuring alignment with semantic intent, deployability, security, and policy compliance remains a more challenging task [19]–[24]. Feedback during iterative deployment, Retrieval-Augmented Generation (RAG), graph-based knowledge integration, constrained decoding, policy enforcement, and multi-agent correction can significantly enhance the technical soundness of solutions. However, these approaches typically assume that an expert has already provided a description of the desired infrastructure. Consequently, they evaluate an LLM's ability to implement a formulated intent, rather than its ability to identify an implicit or missing intent regarding DevOps processes based on the application's source code.

This issue is also closely linked to specification-based development. Specification-based approaches aim to clearly define requirements, architectural structures, behavioral rules, interfaces, constraints, and acceptance criteria prior to or during the creation of artifacts. Recent research has addressed topics such as specification-based code migration, intermediate textual representations, architecture metamodels, code-to-documentation-to-code transformations, and the synthesis of monitoring systems using LLMs [25]–[32]. In particular, the specification-based "code–text–code" reengineering process employs a neutral textual representation to monitor semantic drift during software evolution [25], whereas a unified architecture metamodel introduces structured architectural views as an interface between the developer's intent and generative models [26]. These studies indicate that an explicit intermediate specification can improve the traceability, reproducibility, and control of transformation processes [25], [26].

This study evaluates repository-driven generation of Dockerfiles and Docker Compose configurations for three heterogeneous multi-service applications. The experiment focuses on the distinction between end-to-end functional correctness and conformity with deployment intent.

This distinction motivates separating functional feasibility from fitness for purpose. A generated environment may successfully build and run, but may not replicate the developer's desired security boundaries, production architecture, build workflow, portability constraints, or maintenance practices. In other words, functional correctness and deployment-intent fidelity represent distinct evaluation dimensions.

The identified distinction substantiates the concept of Specification-Driven DevOps. In this study, this approach is viewed as a managed transformation process in which directly observable deployment facts are derived from repository artifacts, while DevOps decisions that are not directly observable or are insufficiently defined are specified via a compact and explicit specification. Thus, the repository is not regarded as an exhaustive specification; instead, it becomes one of the information sources within a hybrid generation model:

$$E = G(R, S_{min}, K, M), \quad (1)$$

where $E$ is the generated deployment environment, $R$ is the software repository, $S_{min}$ is a minimal explicit deployment specification, $K$ denotes external technological knowledge, and $M$ is the generative model.

The proposed approach does not attempt to duplicate in a specification all information already available in source code. Its purpose is to identify the irreducible specification layer:

$$S_{min} = S_{required} - S_{reliably\ inferred}. \quad (2)$$

Repository artifacts may provide evidence about runtime topology and dependencies, whereas workflow, security, platform, and production policies may require explicit specification.

Accordingly, the objective of this study is to investigate repository-driven generation of multi-service deployment environments, formalize the boundary between inferable runtime knowledge and explicit DevOps intent, and derive a minimal complementary deployment specification.

To achieve this objective, the following research tasks are addressed:

- To formalize the transformation of a software repository into a multi-service deployment environment, including the input artifact space, generated Dockerfiles, Compose configurations, services, networks, volumes, and secrets.
- To develop a taxonomy of DevOps knowledge according to its inferability from repository artifacts, distinguishing directly observable facts, derivable facts, underdetermined values, and intent-only constraints.
- To formalize the DevOps intent gap between functionally executable generated environments and environments that conform to architectural, security, workflow, and operational requirements.
- To define an executable validation oracle for assessing the complete multi-service pipeline rather than relying only on syntax, compilation, image construction, or textual similarity.
- To derive a minimal explicit deployment specification containing the information that cannot be inferred reliably from repository artifacts.
- To define an evaluation framework distinguishing build success, runtime functionality, structural completeness, traceability, and deployment-intent fidelity.

The scientific contribution of this work consists of formalizing the boundary between repository-derived DevOps knowledge and explicitly specified deployment intent, introducing an execution-oriented evaluation model, and deriving a compact specification layer for multi-service environment generation.

## II. Related Work

### A. LLM-based software generation and executable correctness

Research in neural code generation initially focused on creating code snippets from natural language descriptions, function auto-completion, and solving standard programming benchmark tasks [1], [2], [33]–[36]. LLMs trained on source code demonstrated significant success on HumanEval, MBPP, and similar benchmarks, facilitating the broader adoption of LLMs in software development [1], [33], [34]. Subsequent research expanded the scope of evaluation to include tasks such as repository-level operations, issue resolution, code repair, test generation, software migration, and agent-based software development [4]–[8], [37]–[41].

A persistent challenge in evaluating code generation is the discrepancy between surface-level similarity and execution correctness. Metrics such as BLEU, exact match, CodeBLEU, and embedding similarity can assign high scores to outputs resembling the reference code without confirming the functional correctness of the generated implementation [35], [36], [42]. Consequently, execution-based metrics—such as pass@k, successful test execution, compilation, behavioral equivalence, and end-to-end scenario execution—are increasingly regarded as more reliable criteria for validity [1], [33], [37], [38].

Repository-level generation introduces additional hidden dependencies, project conventions, configuration requirements, execution services, and cross-file relationships. SWE-bench and related benchmarks demonstrate that resolving real-world software issues requires not only local code generation but also repository navigation, test interpretation, modification of multiple artifacts, and verification of changes within execution environments [37], [38]. This has driven the development of tool-using and agentic systems that iteratively inspect source code trees, execute commands, read error messages, modify files, and re-verify [13]–[17], [39]–[41].

This study adopts an execution-oriented approach at the level of complete multi-service environment generation. Instead of determining whether an LLM can generate a specific function or solve a problem within a repository, the study asks whether the data available in the repository is sufficient to deploy an environment in which a multi-service application operates. Consequently, the primary criterion for correctness is an end-to-end functional oracle that verifies the operation of the deployed service pipeline.

### B. *Automated Dockerfile and environment generation*

Automated environment creation emerged even before the advent of modern systems based on LLMs. Early tools attempted to determine environment dependencies by analyzing source code imports, package repositories, build files, and patterns detected in Dockerfiles [9]–[12]. DockerizeMe employed an environment-dependency inference mechanism to create execution environments for code snippets, whereas other approaches involved analyzing repositories or examining the relationships between project characteristics and container configurations [9], [10]. These systems established the feasibility of deriving parts of an execution environment from project evidence, but their applicability was constrained by programming language, ecosystem, or dependency representation.

Rosa et al. formulated Dockerfile generation as a sequence-generation problem and trained a T5 model using structured Dockerfile requirements extracted from a corpus of 670,982 Dockerfiles [11]. Their results showed that neural generation did not clearly outperform information-retrieval baselines, emphasizing the difficulty of representing environment requirements and validating generated configurations. Importantly, the model input was an explicit structured description derived from Dockerfiles, not raw application source code.

Repo2Run represents a significant step toward the synthesis of repository-based environments [13]. The system takes a Python repository as input, creates an isolated configuration environment, builds a Docker image, runs tests, analyzes failures, and synthesizes a final Dockerfile based on successful commands. A build and test success rate of 86% demonstrates that the feedback obtained during iterative execution enables the resolution of issues related to incomplete dependency declarations and environment inconsistencies. Nevertheless, its target remains a single Docker image capable of running repository tests, not a complete distributed environment with multiple application and infrastructure services [13].

RepoLaunch generalizes automated repository build and test management across programming languages and operating systems [14]. SWE-Builder, introduced within SWE-Factory, distributes repository analysis, environment construction, error interpretation, and validation among specialized agents [15]. Multi-Docker-Eval introduces multilingual repository tasks and demonstrates that environment setup remains challenging outside of homogeneous Python environments [16]. HerAgent proposes a hierarchical model of environment maturity and argues that dependency installation or partial test execution serves as a weaker oracle than the successful execution of the repository's core functionality [17].

These studies establish three important principles. First, the environment creation process should be evaluated at the execution stage. Second, feedback obtained during code compilation, building, testing, and execution helps improve the creation process. Third, creating an environment at the repository level is a significantly more complex task than local code generation.

However, a methodological gap remains: the primary unit of generation in these studies is an individual, test-oriented container. An environment is considered successfully configured if all dependencies are installed and the existing test suite or main entry point can be executed. In contrast, a system composed of multiple services requires orchestration of frontend applications, backend services, background processes, databases, message brokers, secrets, shared networks, health check mechanisms, persistent storage systems, and external access policies. The generated artifacts must be internally consistent across the various Dockerfiles and the Compose model. This study directly addresses the aforementioned problem.

### C. *Infrastructure as Code generation*

IaC research provides a broader perspective on the relationship between syntactic validity, deployability, and intent alignment. IaC artifacts define cloud resources, networking, access controls, storage, policies, and deployment dependencies in machine-processable formats. LLM-based generation of Terraform, CloudFormation, and AWS CDK therefore resembles multi-service Compose generation in that the output must satisfy external platform semantics rather than only language syntax.

IaC-Eval introduced hundreds of human-authored Terraform scenarios and demonstrated that LLM performance for infrastructure programs was significantly lower than performance on standard Python generation benchmarks [19]. Many results were syntactically correct but semantically incompatible with the required infrastructure behavior. Subsequently, IaCGen employed iterative validation and real-time deployment feedback to improve deployment success rates, while revealing that adherence to user intent and security remained far lower than technical deployability [20].

Knowledge-injection approaches further demonstrate that retrieval and structured context can improve technical validity [21]. Graph-based RAG, platform documentation, error taxonomies, and cloud emulation can help models choose valid resource properties and correct API usage. Multi-IaC-Eval extends assessment across Terraform, CloudFormation, and CDK and confirms that syntax is no longer the sole or even primary bottleneck for frontier models [22]. Multi-agent and policy-aware systems additionally employ constrained decoding, orchestration, and Open Policy Agent rules to improve reliability [23], [24].

Prior IaC studies distinguish technical deployability from semantic, policy, and security conformity. This distinction motivates the separate evaluation of functional correctness and deployment-intent fidelity in the present study. The key difference is the direction of knowledge acquisition. Existing IaC benchmarks ordinarily provide an expert-authored natural-language infrastructure request. They test transformation from explicit infrastructure intent into code:

$$S_{IaC} \rightarrow C_{IaC}. \quad (3)$$

The present study begins with application repository evidence and asks which parts of the intended deployment model can be reconstructed:

$$R \rightarrow \hat{S}_{DevOps} \rightarrow C_{Compose}. \quad (4)$$

Thus, the main uncertainty lies not only in code generation but also in specification recovery.

### D. Empirical studies of Docker Compose and multi-container systems

Studies focusing on Docker and multi-container composition cover aspects such as dependency management, orchestration patterns, maintainability, security, code smells, and configuration evolution [43]–[48]. Empirical analysis of Docker Compose repositories has identified typical service structures, database combinations, dependency patterns, networks, volumes, environment variables, and orchestration practices [43]. Such studies demonstrate how developers actually describe multi-container systems and establish an important structural foundation for generated environments.

Studies on the quality of Dockerfiles and Compose files have also revealed a number of common issues, including excessive image sizes, failure to pin dependency versions, inefficient caching, excessive privileges, secret data leakage, outdated base images, a lack of health checks, and improper port exposure [44]–[48]. These findings indicate that a configuration may be executable yet fail to meet requirements regarding maintainability, security, portability, or production readiness.

This literature supports a multidimensional evaluation model. A generated environment should not be assessed solely through “*docker compose up*” or an HTTP success response. Its quality also depends on:

- service and dependency completeness;
- internal consistency;
- minimal host exposure;
- secret-management correctness;
- network isolation;
- deterministic image versions;
- build efficiency;
- health and restart behavior;
- production serving;
- platform compatibility;
- resource constraints.

Together, these dimensions motivate an evaluation model that distinguishes executable correctness from structural, security, workflow, portability, and production-related properties.

### E. Specification-driven development and intermediate representations

Specification-driven development treats specifications as active engineering assets rather than static documentation. Formal and semi-formal specifications can define requirements, interfaces, state transitions, data models, constraints, architecture, deployment policies, and acceptance criteria. Model-based engineering, domain-specific languages, contract-first development, and executable specifications share a common goal: to reduce the semantic gap between intent and implementation [49]–[53].

In the context of development involving LLMs, specifications have taken on particular importance, as natural language prompts are often incomplete, ephemeral, and difficult to verify. Direct code-to-code conversion may preserve the external structure while simultaneously introducing semantic drift, hidden behavioral changes, erroneous assumptions, or a loss of traceability. Specification-based "code–text–code" re-engineering addresses this problem by introducing a neutral textual specification that describes behavior, identifiers, control flow, side effects, dependencies, and domain-specific intent [25]. This intermediate specification is iteratively verified against the source code prior to target code generation, enabling the measurement of losses caused by the transformation [25].

The unified architecture metamodel similarly proposes a structured set of architectural representations spanning business, system, and developer layers [26]. It treats architectural diagrams and related artifacts as an interface between humans and generative models and supports transformations such as code-to-text-to-code. The study indicates that structured architectural context can improve the stability and repeatability of generated artifacts, although redundant views and context selection remain open problems [26].

A specialized migration framework based on LLMs employs approaches—including the use of specifications, functional feature detection, iterative verification, and expert evaluation—to facilitate the transition from Oracle to PostgreSQL [27]. It demonstrates that successful migration depends not only on the syntactic transformation of code but also on preserving database behavior, interrelationships, and the semantics of functional elements. Research on token optimization and dependency-aware grouping addresses the practical challenge of representing large codebases within limited LLM context windows [28]. These principles are directly relevant to repository-driven DevOps generation because service topology and deployment evidence are distributed across manifests, source files, configuration files, proxy declarations, environment variables, and directory structure [27].

Specification-driven synthesis of intelligent monitoring systems extends the same reasoning to monitoring architecture [29]. Software-state formalization and evolution-aware monitoring emphasize that generated or evolving systems must be evaluated not only at the moment of creation but throughout their life cycle [29]–[32]. The Software Functional State concept represents a system through functional adequacy, performance, stability, risk, security exposure, contextual coherence, and SDLC phase [30]. Multi-drift predictive monitoring and context-aware decision support further show how changes in system state can be detected and used to adapt decision strategies [31], [32].

These studies support the view that deployment generation should be treated as a controlled specification transformation rather than an isolated code-generation task. The output environment becomes part of the evolving software state and must preserve both executable behavior and operational constraints.

### F. Knowledge extraction, architecture recovery, and traceability

Generating an environment from source code requires recovering architectural knowledge from implementation evidence. Architecture-recovery research has long used static dependencies, dynamic traces, clustering, naming, repository structure, build files, and configuration data to reconstruct components and relations [54]–[58]. Modern LLMs add semantic interpretation and cross-artifact reasoning but do not eliminate uncertainty.

A repository may provide strong evidence that a service connects to PostgreSQL, but weak or no evidence regarding:

- whether PostgreSQL should be externally reachable;
- whether the deployment requires separate frontend and backend networks;
- whether the target platform is ARM64 or AMD64;
- whether the frontend must use a development server or an Nginx production stage;
- whether credentials should be generated, mounted as files, or obtained from a vault;
- whether builds require multi-stage optimization.

This suggests a four-level knowledge taxonomy:

- Directly observable knowledge—explicit ports, imports, commands, image-compatible runtimes, connection strings, and manifest dependencies.
- Derivable knowledge—service topology, proxy routes, worker roles, and infrastructure services reconstructed from several artifacts.
- Underdetermined knowledge—values for which several internally consistent alternatives are possible, such as credentials or compatible image versions.
- Intent-only knowledge—deployment and policy decisions that have no reliable implementation-level representation.

Traceability is necessary because each generated environment element should be linked to the repository evidence or explicit specification field that supports it:

$$\tau(e_k) = \langle o_k, r_k \rangle, \quad (5)$$

where $o_k$ denotes the origin type and $r_k$ denotes the supporting repository artifact or specification field. Such traceability allows incorrect inferences to be distinguished from missing information and supports later regeneration when the repository or specification changes.

### G. Functional correctness versus production readiness

In modern studies, successful building, compilation, test execution, or deployment are often cited as the primary results. However, these are necessary but insufficient criteria. A development-oriented environment might expose internal ports, use default credentials, mount source directories, run development servers, and rely on mutable image tags while still passing functional tests.

Successful building, startup, or functional execution is necessary but insufficient for establishing production readiness. Structurally different environments may implement equivalent observable behavior while differing in security, maintainability, portability, or operational intent. This distinction also explains why structural equality with a developer-authored Compose file is not an appropriate primary oracle. Multiple structurally different environments can produce equivalent behavior. Structural differences must instead be classified according to their effect:

- functionally equivalent alternative;
- harmless technical variation;
- maintainability degradation;
- portability degradation;
- security degradation;
- operational defect;
- functional defect.

This classification provides the conceptual basis for separating functional correctness from structural and deployment-intent fidelity in the experimental evaluation.

### H. Research gap

The research literature has established a solid foundation for code generation based on LLMs, repository-aware agents, Dockerfile synthesis, environment creation, IaC generation, architecture recovery, and specification-based transformations. However, five limitations remain.

First, most systems that interact with repositories and environments generate only a single Dockerfile for test execution, rather than a coherent deployment environment for multiple services [13]–[17].

Secondly, research on IaC generation assumes that the desired infrastructure intent is explicitly specified, whereas repository-driven generation must first derive a deployment specification [18]–[23].

Thirdly, existing studies rarely systematically distinguish between facts observable from the repository and unobservable DevOps intentions.

Fourthly, successful execution is often viewed as the ultimate criterion, despite evidence that security, networking, build, production, and workflow policies may remain unsatisfied.

Fifth, the reviewed literature does not provide an established compact specification designed specifically to complement repository-based inference.

The present study addresses these limitations by introducing a specification-driven process for multi-service environment generation, a taxonomy of inferable and non-inferable DevOps knowledge, an executable end-to-end oracle, a formal intent-gap model, a multidimensional quality framework, and a minimal deployment specification derived from observed generation omissions and underdetermined decisions.

## III. Method and Experimental Design

### A. Problem formulation

The first research task is to formalize the transformation of a software repository into a complete multi-service deployment environment. Unlike conventional Dockerfile generation, this transformation must identify not only the runtime requirements of an individual application but also the relationships among application services, infrastructure services, background processes, databases, message brokers, proxy components, credentials, networks, volumes, and externally accessible endpoints.

The experimental basis included three heterogeneous repositories containing application source code, dependency manifests, configuration files, and auxiliary repository artifacts. Existing Dockerfiles and Docker Compose configurations were excluded from the generation input. The LLM was required to generate a complete docker-compose.yml file and all Dockerfiles needed to start the applications. Generated environments were subsequently executed and validated through end-to-end functional requests. Let a software repository be represented as:

$$R = \langle F_{src}, F_{dep}, F_{cfg}, F_{sec}, F_{aux}, H \rangle, \quad (6)$$

where $F_{src}$ is the set of application source-code files; $F_{dep}$ is the set of dependency and build manifests; $F_{cfg}$ is the set of application configuration files; $F_{sec}$ is the set of discoverable secret-related files; $F_{aux}$ is the set of auxiliary artifacts, such as proxy configurations, startup scripts, and seed data; $H$ represents the repository hierarchy and relationships among files.

The repository input deliberately excludes the developer-authored deployment artifacts:

$$F_{Dockerfile} \notin R, \qquad F_{Compose} \notin R. \quad (7)$$

The generated deployment environment is represented:

$$E = \langle S, D, C, N, V, Q, P \rangle, \quad (8)$$

where $S$ is the set of generated services; $D$ is the set of generated Dockerfiles; $C$ is the generated Docker Compose configuration; $N$ is the set of defined networks; $V$ is the set of volumes and mount declarations; $Q$ is the set of secrets and credential propagation rules; $P$ is the set of host-port and internal-port policies.

The transformation performed by an LLM-based generation mechanism M can therefore be expressed as:

$$T_M : R \rightarrow \widehat{E}, \quad (9)$$

where $\widehat{E}$ is the environment inferred from repository evidence.

A successful repository-to-environment transformation should satisfy:

$$Complete(\widehat{E}) \wedge Consistent(\widehat{E}) \wedge Executable(\widehat{E}). \quad (10)$$

Completeness $Complete(\widehat{E})$ means that all services required by the application are represented. Consistency $Consistent(\widehat{E})$ means that service names, ports, environment variables, credentials, dependencies, and network references agree across the generated artifacts. Executability $Executable(\widehat{E})$ means that the environment can be built, started, and used to complete an end-to-end application scenario.

### B. *Transformation pipeline*

The formalized transformation consists of six main stages:

- repository discovery;
- artifact classification;
- service and dependency inference;
- environment-model construction;
- Dockerfile and Compose generation;
- build and functional validation.

The complete repository-to-environment transformation pipeline is shown in Fig. 1, including repository discovery, artifact classification, environment generation, build execution, correction, and functional validation.

The process begins with active repository traversal. The repository-discovery stage traverses both primary source files and auxiliary artifacts that may contain deployment-relevant evidence, including proxy configurations, startup scripts, secret-related files, and initialization resources.

Accordingly, the transformation is modeled as a repository-level evidence-aggregation process in which multiple files jointly determine one deployment decision.

### C. *Input artifact space*

The repository input contains heterogeneous signals. Some signals directly identify deployment properties, whereas others require cross-file interpretation. Table I summarizes the artifact classes, representative files, extracted evidence, and corresponding generated deployment decisions.

TABLE I. FORMALIZED REPOSITORY INPUT ARTIFACT SPACE

| Artifact class | Typical examples | Extracted evidence | Generated deployment decision |
|---|---|---|---|
| Source code $F_{src}$ | `.py`, `.js`, `.cs`, `.rs`, `.java` | Entry points, listening ports, database clients, background workers, hardcoded hostnames | Application services, commands, internal ports, infrastructure dependencies |
| Dependency manifests $F_{dep}$ | `requirements.txt`, `package.json`, `Cargo.toml`, `pom.xml`, `.csproj` | Runtime version, framework, libraries, build system | Base image, dependency installation, build commands |
| Configuration files $F_{cfg}$ | `.env.example`, application.properties, JSON/YAML configuration | Environment-variable names, defaults, database URLs, proxy targets | Compose environment section, hostnames, credentials, service links |
| Secret-related files $F_{sec}$ | `db/password.txt`, key files, secret templates | File-based credential mechanism | Compose secrets declaration and service secret mounts |
| Auxiliary files $F_{aux}$ | `setupProxy.js`, startup scripts, database initialization files | Hidden routes, proxying, initialization procedures | Reverse proxy configuration, service dependency, initialization mounts |
| Repository structure $H$ | Service directories, nested modules, frontend/backend folders | Service boundaries and build contexts | Separate services, Dockerfile locations, Compose build contexts |

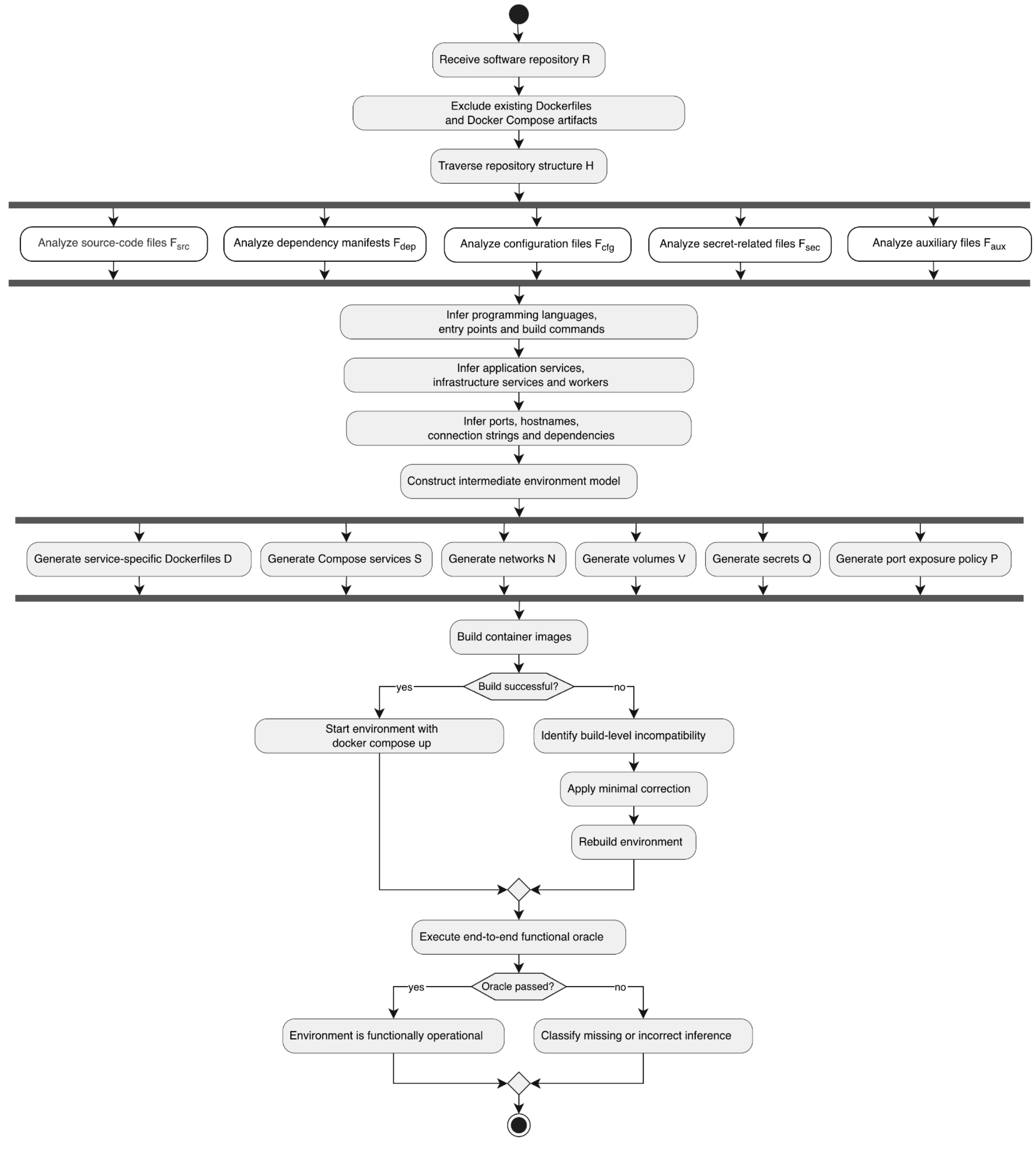

Fig. 1. Repository-to-Multi-Service Deployment Environment Transformation

As illustrated in Fig. 1, repository artifacts are first analyzed and consolidated into an intermediate environment representation before Dockerfiles, Compose configurations, networks, volumes, secrets, and port policies are generated. The relation between repository evidence and generated artifacts can be represented as:

$$\Gamma:\ F_R \rightarrow A_E. \tag{11}$$

where $F_R$ is the set of repository artifacts and $A_E$ is the set of generated environment elements. For example:

*app.run(port=80)→internal application port→expose: 80,*

*new Redis("redis")→Redis dependency→service: redis,*

*db/password.txt→file-based secret→MYSQL_ROOT_PASSWORD_FILE.*

A generated artifact can be supported by more than one repository signal. Therefore, the general mapping is many-to-many:

$$\Gamma \subseteq F_R \times A_E. \tag{12}$$

The mappings summarized in Table I show that a generated deployment element may depend on evidence distributed across source code, manifests, configuration files, auxiliary artifacts, secret-related files, and repository structure.

*D. Intermediate deployment-environment model*

Before Dockerfiles and Compose configurations are generated, repository evidence should be represented as an intermediate environment model:

$$M_E = \langle S_M, R_M, I_M, B_M, Q_M \rangle, \quad (13)$$

where $S_M$ is the inferred service set; $R_M$ is the inter-service relation set; $I_M$ is the infrastructure model; $B_M$ is the build model; $Q_M$ is the credential and secret model.

Each service $si_i \in S_M$ is formalized as:

$$s_i = \langle n_i, t_i, l_i, b_i, c_i, p_i, e_i, d_i, v_i, q_i \rangle, \quad (14)$$

where $n_i$ is the service name; $t_i$ is the service type; $l_i$ is the programming language or infrastructure technology; $b_i$ is the build context; $c_i$ is the runtime command; $p_i$ is the port set; $e_i$ is the environment-variable set; $d_i$ is the dependency set; $v_i$ is the volume-mount set; $q_i$ is the secret set.

The service type is selected from:

$$t_i \in \{frontend, backend, worker, database, broker, proxy, auxiliary\}. \quad (15)$$

Relations between services are represented as a directed graph:

$$G_E = (S_M, R_M), \quad (16)$$

where an edge $(s_i, s_j, r_{ij}) \in R_M$ means that service $s_i$ depends on, connects to, proxies to, reads from, or writes to service $s_j$.

The structure of the intermediate deployment-environment model and its relation to repository evidence and generated artifacts are illustrated in Fig. 2.

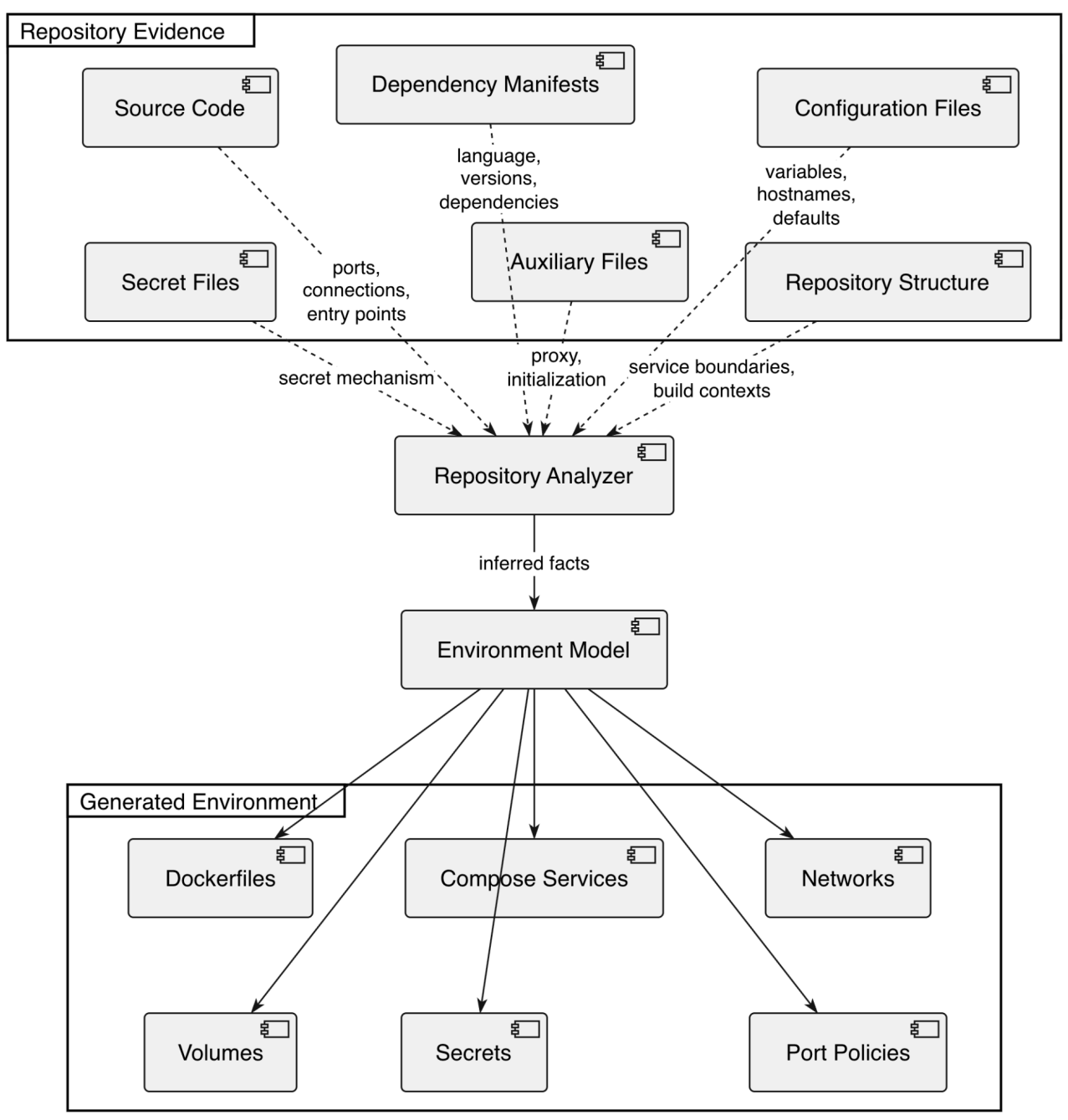

Fig. 2. Intermediate environment model generated from repository evidence

As shown in Fig. 2, the intermediate model consolidates repository-derived evidence before the deployment artifacts are serialized. This separation allows the same inferred model to support multiple representations and validation activities. The same inferred model can be rendered as:

- Dockerfiles;
- `docker-compose.yml`;
- a deployment diagram;
- a service-dependency graph;
- a traceability table;
- a validation plan.

The intermediate model preserves the origin of each generated decision; the proposed traceability model is defined in Section VI.I.

*E. Experimental configuration*

The experiment used GPT Codex 5.3 through GitHub Copilot. Generated environments were evaluated using deterministic end-to-end HTTP oracles and manual structural comparison against developer-authored deployment artifacts. Execution was performed on an Apple M1 Pro host with 32 GB RAM on the ARM64 platform. The complete experimental configuration is summarized in Table II.

TABLE II. EXPERIMENTAL CONFIGURATION

| Component | Details |
|---|---|
| LLM model | GPT Codex 5.3 via GitHub Copilot |
| Evaluation method | Functional end-to-end oracle using an HTTP request through the complete service pipeline |
| Ground-truth comparison | Manual structural comparison with developer-authored Dockerfiles and Compose configurations |
| Repositories tested | 3 |
| Host platform | Apple M1 Pro, 32 GB RAM, ARM64 |

Under the configuration summarized in Table II, the repositories were selected to cover different language combinations, service topologies, and credential-storage patterns.

*F. Subject repositories*

Each application included the source artifacts required for local execution, used locally built application images, supported startup through a single Compose command in its reference implementation, and was executed on the ARM64 host used in the experiment.

The first system was selected to evaluate heterogeneous service inference and background-worker identification. The second evaluated generation when credential values were unavailable in repository input. The third evaluated whether a standalone password file could be interpreted as evidence of a Docker secrets mechanism. The technology stacks, services, credential patterns, and principal repository signals of the three systems are summarized in Table III.

TABLE III. CHARACTERISTICS OF THE EVALUATED REPOSITORIES

| Repository | Technology stack | Services | Credential pattern | Notable repository signal |
|---|---|---|---|---|
| `example-voting-app` | Python, Node.js, C#/.NET | `vote, result, worker, Redis, PostgreSQL` | Database credentials hardcoded in `Program.cs` | Three programming languages, asynchronous worker, queue-based communication |
| `react-rust-postgres` | React, Rust with Actix Web, PostgreSQL | `frontend, backend, db` | Credentials present only in the reference Compose file and therefore unavailable to the LLM | Hidden http-proxy-middleware configuration; application port ambiguity |
| `react-java-mysql` | React, Java with Spring Boot, MariaDB/MySQL | `frontend, backend, db` | Password stored in `db/password.txt` | Docker-secret inference and Maven-based build structure |

The repository characteristics presented in Table III provide the experimental basis for evaluating topology inference, credential propagation, auxiliary-file discovery, and deployment-intent omissions.

### G. Service-Topology Inference

Repository evidence is transformed into a directed service-dependency graph $G_E = (S_M, R_M)$, where the nodes represent application, infrastructure, worker, and auxiliary services, and the edges represent communication, proxying, data access, or processing dependencies. Repository-specific reconstruction results are presented in Section V.

### H. Dockerfile generation

For every application service built from source, the transformation generated a service-specific Dockerfile:

$$D = \{d_1, d_2, ..., d_m\}, \tag{17}$$

where $m$ is the number of application services requiring local builds.

The generation of a Dockerfile $d_i$ includes the inference:

$$d_i = \langle I_i, W_i, X_i, K_i, C_i \rangle, \tag{18}$$

where $I_i$ is the base image; $W_i$ is the working directory; $X_i$ is the file-copy strategy; $K_i$ is the dependency installation and build sequence; $C_i$ is the runtime command.

Generated Dockerfiles were evaluated according to base-image compatibility, dependency installation, build commands, runtime commands, build-stage structure, caching strategy, production-serving mode, and platform support. The repository-specific results are reported in Sections IV and V.

### I. Compose configuration generation

The generated Compose model can be represented as:

$$C = \langle S_C, N_C, V_C, Q_C \rangle, \tag{19}$$

where $S_C$ contains service definitions; $N_C$ contains named networks; $V_C$ contains volumes; $Q_C$ contains secrets.

Each generated service definition includes:

$$c_i = \langle build, image, command, ports, environment, \\ depends_on, networks, volumes, secrets \rangle. \tag{20}$$

The generated Compose configurations were evaluated for service completeness, hostname consistency, dependency ordering, infrastructure integration, network declarations, volume definitions, secret propagation, and host-port exposure. Detailed results are presented in Section IV.

### J. Volume and mount generation

Volumes can serve several distinct purposes:

$$V = V_{persistent} \cup V_{development} \cup V_{initialization} \cup V_{secret}. \tag{21}$$

These categories correspond to:

- persistent database storage;
- source-code mounts for live reload;
- initialization scripts or seed data;
- file-based secret mounts.

A source repository can reveal that an application supports a development server, but it does not necessarily state that source code should be mounted into the container. Therefore:

$$development\ mount\ policy \not\subseteq \\ reliably\ observable\ repository\ facts. \tag{22}$$

The transformation distinguishes persistent, development, initialization, and secret-related mounts. Whether a particular mount should be generated depends on both repository evidence and the declared deployment mode.

### K. Secret and credential generation

Credential handling was classified into three cases observed in the experiment: directly available values, required but unavailable values, and repository-visible file-based secret mechanisms. An additional policy-driven case was introduced analytically for situations in which organizational secret-management requirements override repository defaults.

### L. Functional validation of the transformation

The generated environment was not considered correct merely because the Compose file parsed or the container images were built. Each application was validated by an end-to-end functional oracle. Table IV presents the oracle command and expected result for each evaluated repository.

TABLE IV. END-TO-END FUNCTIONAL ORACLES

<table>
<tr><th>Application</th><th>Oracle command</th><th>Expected result</th></tr>
<tr><td>example-voting-app</td><td>curl -X POST localhost:8080 -d 'vote=a' && curl localhost:8081 | grep '"a":[0-9]*'</td><td>The submitted vote is propagated through Flask → Redis → .NET worker → PostgreSQL → Node.js and appears in the result output.</td></tr>
<tr><td>react-rust-postgres</td><td>curl localhost:3000/api/users</td><td>[], confirming React proxy → Rust backend → PostgreSQL</td></tr>
<tr><td>react-java-mysql</td><td>curl localhost:3000/api</td><td>{"id":1,"name":"Docker"}, confirming React → Spring Boot → MariaDB/MySQL</td></tr>
</table>

The defined oracles were applied after the generated images had been built and the complete Compose environments had been started. The sequence of build, startup, request execution, and expected-response verification is illustrated in Fig. 3.

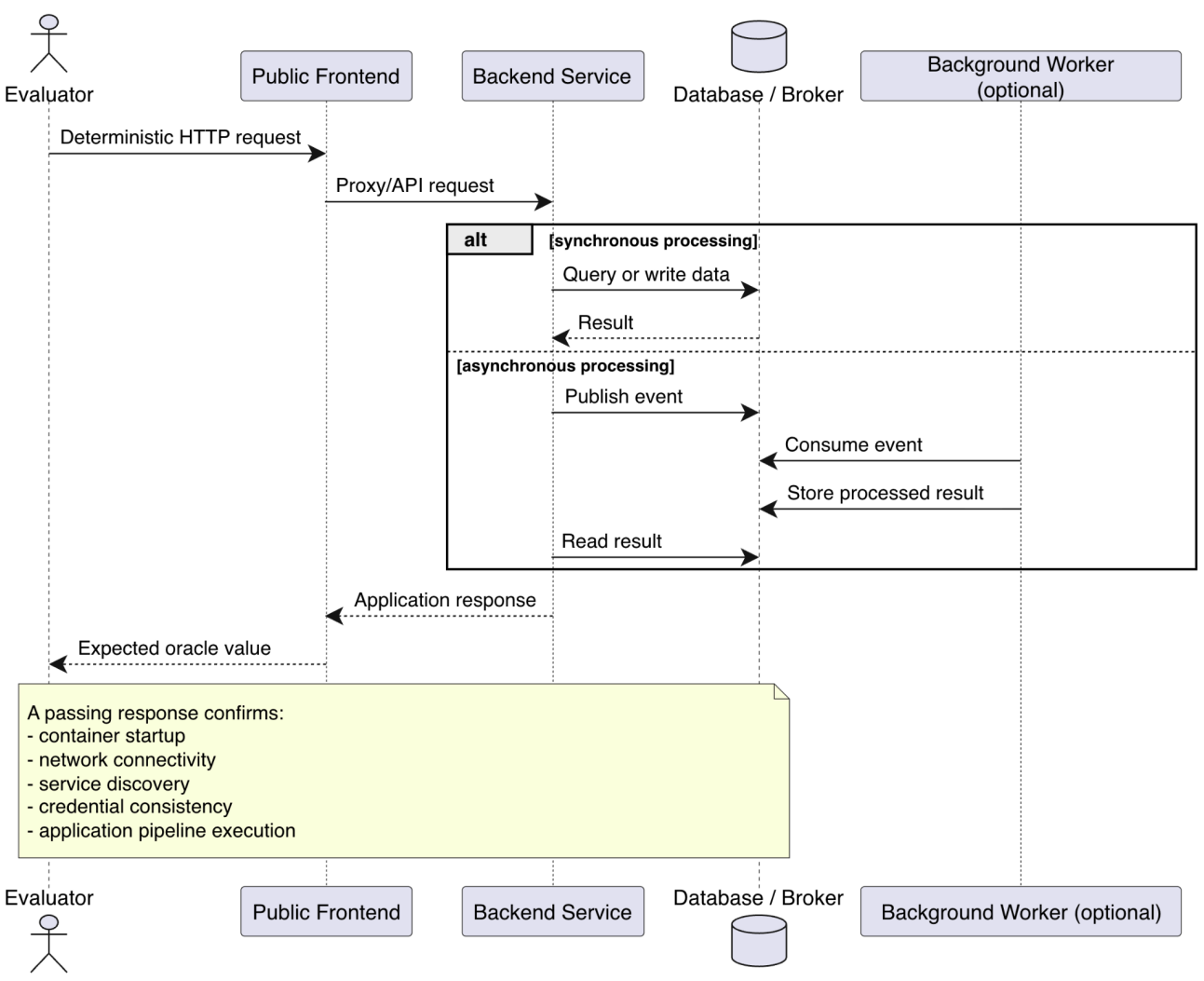

Fig. 3. Functional Oracle for a Generated Multi-Service Environment

Together, Table IV and Fig. 3 define the functional-validation procedure used to obtain and interpret the experimental results presented in the following section.

## IV. Results and Discussion

### A. Evaluation perspective

The generated deployment environments were evaluated along two complementary dimensions:

- Functional correctness, defined as the ability to build, start, and execute the complete multi-service application pipeline.
- Deployment-intent fidelity, defined as the degree to which the generated Dockerfiles and Compose configuration reproduced the developer's architectural, security, workflow, and production decisions.

These dimensions must be assessed separately because a generated environment can be operational without being structurally identical to the developer-authored deployment configuration. The experiment therefore used an end-to-end functional oracle as the primary correctness criterion and a manual structural comparison with the developer artifacts as a secondary analytical baseline. The oracle was deliberately designed to validate observable behavior rather than textual or structural similarity.

The generated and developer-authored environments may differ structurally while implementing the same tested behavior; therefore, strict artifact identity is not used as the primary correctness criterion.

Functional equivalence is denoted by:

$$E_G \equiv_F E_D, \tag{23}$$

while structural similarity is denoted by:

$$Sim_S(E_G, E_D). \tag{24}$$

Accordingly, the experiment considered a generated environment successful when:

$$Build(E_G) \wedge Start(E_G) \wedge Oracle(E_G) = Pass. \tag{25}$$

Structural differences were then analyzed to determine whether they represented:

- acceptable alternative implementations;
- internal-consistency-preserving substitutions;
- missing workflow intent;
- production-readiness degradation;
- security-relevant divergence;
- functional errors.

This classification was applied after functional validation to distinguish executable alternatives from genuine functional and deployment-intent failures.

### B. Functional correctness of the generated environments

All three generated multi-service environments became operational and passed their corresponding end-to-end oracle.

The final operational success rate was:

$$SR_{operational} = \frac{3}{3} = 1.00. \tag{26}$$

A stricter first-build success rate was:

$$SR_{first-build} = \frac{2}{3} \approx 0.667. \tag{27}$$

Repository-level startup, build, oracle, and final operational results are detailed in Table V.

TABLE V. Functional Results of Repository-Driven Environment Generation

| Repository | Compose startup | Build-level issue | Functional oracle | Final result |
|---|---|---|---|---|
| `example-voting-app` | Pass | None | Pass | Operational |
| `react-rust-postgres` | Pass after base-image correction | Rust 1.85 was incompatible with the current Actix Web dependency; updated to Rust 1.88 | Pass | Operational |
| `react-java-mysql` | Pass | None | Pass | Operational |

As shown in Table V, two environments succeeded on the first build, while the Rust environment became operational after a base-image correction. The distinction is important because the correction did not alter the inferred service topology, application ports, database dependency, proxy relationship, or Compose orchestration.

The Rust case shows that topology inference and runtime-version selection should be evaluated separately.

### C. End-to-end oracle results

Applying the oracles defined in Table IV confirmed successful execution of the complete service path for each application rather than the liveness of isolated containers.

The empty array returned by react-rust-postgres was a valid result rather than evidence of missing functionality. The database contained no seed data; therefore, [] confirmed that the frontend proxy, backend process, database connection, service discovery, and response path were operating correctly.

A passing oracle provided joint evidence of successful image construction, container startup, inter-service communication, and execution of the tested application path.

However, the oracle did not validate all aspects of production quality. It did not test authentication, error handling, persistence under restart, performance, load behavior, recovery, or security policy. Functional success should therefore be interpreted as basic end-to-end operability, not complete production readiness.

*D. Accuracy of repository-derived signal inference*

A central result was the successful inference of runtime-relevant information. Table VI summarizes the observed reconstruction rates for language ecosystems, base-image compatibility, application ports, infrastructure services, hostnames, depends_on relationships, background workers, Docker secrets, and hidden proxy configurations. The evidence was distributed across source code, dependency manifests, framework configuration, connection strings, startup logic, proxy declarations, filenames, and repository structure. Therefore, deployment inference was based on both direct extraction and cross-artifact reasoning.

TABLE VI. INFERENCE RESULTS FOR OBSERVABLE DEPLOYMENT SIGNALS

| Signal category | Result |
|---|---|
| Correct language ecosystem | 3/3 |
| Initial base-image compatibility | 2/3 |
| Correct application ports | 3/3 |
| All required infrastructure services | 3/3 |
| Consistent infrastructure hostnames | 3/3 |
| Correct `depends_on` relationships | 3/3 |
| Background worker identification | 1/1 |
| Docker secrets pattern inference | 1/1 |
| Hidden proxy configuration inference | 2/2 |

This can be represented as:

$$a_j = g(f_1, f_2, ..., f_k), \tag{28}$$

where $a_j$ is a generated deployment decision and $f_1, ..., f_k$ are repository-level evidence items.

The Docker secrets mechanism required cross-artifact reasoning over db/password.txt, the database technology, and container conventions.

*E. File discovery and non-obvious evidence*

The model identified two non-obvious auxiliary artifacts: a frontend proxy configuration and a file-based database secret. Their repository-specific mappings are analyzed in Section V.

This observation supports the following repository-analysis model:

$$R \rightarrow Discover(R) \rightarrow Classify(F) \rightarrow Infer(A_E). \tag{29}$$

Repository-wide traversal enabled identification of the auxiliary artifacts observed in the experiment.

Nevertheless, these findings should be interpreted cautiously. Proxy inference was observed in two applicable repositories, while Docker-secret inference was observed in one repository. They demonstrate capability but do not yet establish broad statistical reliability.

*F. Internal consistency versus developer-ground-truth alignment*

One of the most important findings arose in `react-rust-postgres`. The original database credentials existed only in the excluded developer-authored Compose configuration. They were therefore unavailable to the model.

The model generated alternative credentials but used them consistently in:

- the PostgreSQL service;
- the Rust backend environment;
- the generated connection configuration.

Thus:

$$q_{generated,db} = q_{generated,backend}, \tag{30}$$

while:

$$q_{generated} \neq q_{developer}. \tag{31}$$

Despite this inequality, the environment passed the functional oracle.

This means:

$$E_G \neq E_D, \tag{32}$$

but:

$$E_G \equiv_F E_D. \tag{33}$$

For underdetermined values, internal consistency within the generated environment may be more relevant to the tested behavior than literal reproduction of unavailable developer values.

The observed relationship can be formalized as:

$$OraclePass(E_G) \Rightarrow$$

$$InternalConsistencytested\,path(E_G). \tag{34}$$

Structural comparison distinguished valid alternatives, consistent underdetermined values, version drift, omitted workflow, security, and production intent, and functional defects. Table VII summarizes these different classes, representative examples, and their effects.

TABLE VII. INTERPRETATION OF GENERATED-VERSUS-DEVELOPER DIFFERENCES

| Difference type | Example | Effect |
|---|---|---|
| Functionally equivalent alternative | Generated database password differs but is propagated consistently | Functionally acceptable in the evaluated environment |
| Compatible technical alternative | Java 11 instead of Java 17, application remains compatible | Functionally compatible in the tested environment |
| Version drift | Rust 1.85 instead of required Rust 1.88 | Build failure until corrected |
| Missing workflow intent | No live-reload mounts or multi-stage builds | Functionally operational but inconsistent with the developer workflow |
| Missing security intent | Backend port exposed to host | Functional but less restrictive |
| Missing production intent | Development server used instead of Nginx | Functional but not production-ready |

The classifications in Table VII show why artifact differences cannot be treated uniformly: some preserve the tested behavior, whereas others represent build failures or deployment-intent degradation. Accordingly, the comparison distinguishes functional equivalence $E_G \equiv_F E_D$ from structural fidelity $Fid_S(E_G, E_D)$.

### G. Systematic omissions

Although runtime-relevant signals were inferred successfully, several categories were consistently absent from the generated environments.

Table VIII summarizes the systematic omissions observed across the three repositories. These omissions primarily concern workflow, security, build optimization, portability, and production policy rather than the runtime path exercised by the functional oracle.

TABLE VIII. SYSTEMATIC OMISSIONS ACROSS THE EVALUATED REPOSITORIES

| Category | Result | Interpretation |
|---|---|---|
| Network segmentation | 0/3 | Flat network generated in every case |
| Multi-stage Dockerfiles | 0/3 | No separate development, build, and final stages |
| Explicit dependency-layer caching | 0/3 | No optimized copy/install sequence matching developer strategy |
| Live-reload volume mounts | 0/3 | Development workflow not reconstructed |
| Production frontend serving | 0/2 | Development server retained instead of Nginx |
| Correct restrictive port-exposure policy | 0/2 applicable cases | Internal backend ports mapped to the host |
| Cross-platform build strategy | 0/3 | No explicit `$BUILDPLATFORM` or `$TARGETARCH` logic |

The results in Table VIII reveal that the recurring deficiencies were concentrated in non-functional deployment decisions rather than in service discovery or runtime connectivity. The most systematic security-related omission concerned network segmentation.

### H. Network segmentation and security intent

The repeated absence of network segmentation indicates that service connectivity was easier to infer than intended trust boundaries.

A flat network can be represented as:

$$N_G = \{n_{default}\}. \tag{35}$$

where all services are mutually reachable.

A segmented environment might instead require:

$$N_D = \{n_{public}, n_{application}, n_{data}\}. \tag{36}$$

Application source code can indicate that service $A$ connects to service $B$, but it usually cannot indicate that service $C$ must be isolated from service $A$. This is a negative architectural constraint rather than an application-level dependency.

Therefore:

$$Connects(A, B) \tag{37}$$

may be observable, while:

$$\neg Reachable(A, C) \tag{38}$$

is usually not.

Network-isolation policy therefore belongs to explicit deployment intent.

### I. Port over-exposure

The model over-exposed backend ports in both applicable frontend–backend systems.

In `react-rust-postgres`, it added:

```
ports:
  - "8000:8000"
```

although the developer intended the backend to remain internal.

In `react-java-mysql`, it similarly exposed port 8080, although only the frontend was intended to be publicly reachable. The voting application did not exhibit this divergence because both browser-facing services were intentionally exposed.

This difference is security-relevant because unnecessary host exposure expands the reachable attack surface.

The generated behavior reflects a plausible model default:

$$known\ application\ port \rightarrow host\ port\ mapping.$$

However, the correct production rule is often:

$$ListenPort(s, p) \nRightarrow PublishHostPort(s, p).$$

A repository may reveal a listening port but not whether it should be published, kept internal, routed through a proxy, or enabled only for development. Port exposure therefore belongs to explicit DevOps intent.

### J. Build-strategy omissions

The generated Dockerfiles were sufficient for execution but did not reproduce the developer-authored multi-stage or dependency-layer caching strategies.

A build command may be observable in a manifest, but the intended development or production build strategy remains underdetermined.

The omission is interpreted as missing deployment intent in the generation input rather than as evidence of insufficient general Docker knowledge.

### K. Development versus production behavior

Live-reload mounts were absent in all three generated environments, while production `Nginx` serving was absent in both applicable frontend applications.

These two omissions point in different directions:

- no live-reload mounts means the environment was not optimized as a complete development workflow;
- no Nginx production stage means the frontend was also not production-oriented.

The generated configurations were runnable but did not clearly implement either a complete development workflow or a production-oriented deployment. This occurred because the requested target mode was unspecified:

$$EnvironmentMode \in \{dev, test, staging, prod\}$$

was not available in repository evidence.

Without this variable, several deployment decisions remain underdetermined.

### L. Secret inference and secret-policy limits

The `react-java-mysql` case demonstrated successful inference of a file-based secret mechanism. The model discovered the password file, declared it as a Compose secret, mounted it into the required services, and used the _FILE form of the database environment variable. All four inference steps were correct.

The observed secret-mechanism inference does not establish general secret-policy inference.

The experiment distinguishes:

$$secret\ mechanism\ visible\ in\ repository$$

from:

$$organization - mandated\ secret\ policy.$$

A file such as `db/password.txt` provides positive evidence for a file-based mechanism. By contrast, source code alone cannot indicate whether an organization requires:

- .env files;
- Docker secrets;
- Kubernetes secrets;
- HashiCorp Vault;
- cloud-native secret stores;
- runtime-generated credentials.

The file-based mechanism was inferable in the observed case, whereas organizational secret policy remains an explicit specification concern.

### M. Cross-repository knowledge boundary

The cross-repository results define a practical boundary between information that can be inferred from repository artifacts and information that should be explicitly specified. This boundary is summarized in Table IX using the observable, derivable, underdetermined, and intent-only knowledge classes.

TABLE IX. EMPIRICAL BOUNDARY BETWEEN INFERABLE AND SPECIFICATION-REQUIRED KNOWLEDGE

| DevOps Knowledge | Knowledge class | Explicit specification required? |
|---|---|---|
| Service topology | Observable/derivable | Usually no |
| Application ports | Observable | Usually no |
| Infrastructure dependencies | Observable/derivable | Usually no |
| Background workers | Derivable | Usually no |
| Hidden proxy configuration | Derivable | Usually no |
| Missing credentials | Underdetermined | Yes, when exact policy matters |
| Network segmentation | Intent-only | Yes |
| Public/internal classification | Intent-only | Yes |
| Build strategy | Intent-only | Yes |
| Production serving | Intent-only | Yes |
| Runtime version pinning | Underdetermined | Yes, when fixed versions matter |

As shown in Table IX, DevOps knowledge is classified as observable, derivable, underdetermined, or intent-only. Observable and derivable knowledge was generally reconstructable in the evaluated repositories, whereas underdetermined and intent-only knowledge required explicit specification or controlled defaults.

### N. Functional equivalence and structural fidelity

Functional equivalence and structural fidelity are distinct evaluation dimensions.

Let:

$$F(E_G) \tag{39}$$

denote the behavior of the generated environment, and:

$$S(E_G) \tag{40}$$

denote its deployment structure.

Functional equivalence is:

$$E_G \equiv_F E_D \Leftrightarrow F(E_G) = F(E_D) \tag{41}$$

for the evaluated oracle set.

A possible future quantitative structural-fidelity measure is:

$$Fid_S(E_G, E_D) = \frac{\sum_{j=1}^{m} w_j \delta_j}{\sum_{j=1}^{m} w_j}, \tag{42}$$

where $m$ is the number of compared structural properties; $w_j$ is the importance weight of property $j$; $\delta_j = 1$ when the generated and developer environments agree for that property, otherwise $\delta_j = 0$. This weighted measure was not calculated in the conducted experiment.

After the Rust version correction, all three systems achieved functional equivalence under the tested oracles without full structural fidelity. Their qualitative position within the functional-correctness and structural-fidelity space is shown in Table X.

TABLE X. A TWO-AXIS EVALUATION MODEL

| | **High structural fidelity** | **Low structural fidelity** |
|---|---|---|
| **High functional correctness** | Intended and executable environment | Executable alternative or intent-incomplete environment |
| **Low functional correctness** | Structurally similar but broken environment | Failed generation |

As indicated in Table X, the tested systems were qualitatively positioned in the upper-right category: high functional correctness with incomplete structural and deployment-intent fidelity.

### O. Main finding: the signal gap

The dominant limitation was incomplete recovery of deployment intent rather than incorrect reconstruction of runtime dependencies. Therefore, the observed gap can be formalized as:

$$G_{signal} = K_{required} - K_{repository}. \tag{43}$$

The consistently omitted categories belonged to $G_{signal}$:

$$G_{signal} = \{environment\ mode,\ network\ boundaries,\ port\ exposure\ policy,\ build\ strategy,\ production\ serving,\ platform\ constraints,\ resource\ policies\}. \tag{44}$$

The signal gap comprises required deployment decisions that are neither observable nor derivable from repository artifacts.

*P. Implications for Specification-Driven DevOps*

The observed information gaps motivate the following unvalidated hybrid generation model:

$$E_G = G(R, S_{min}), \quad (45)$$

where $R$ provides observable and derivable runtime facts; $S_{min}$ provides non-observable deployment intent.

This model was analytically derived from the observed omissions and was not evaluated through specification-augmented regeneration. The minimal specification covers only deployment decisions that were absent or underdetermined in repository evidence. The relationship between repository-derived facts, explicitly specified intent, and the generated environment is summarized in Fig. 4.

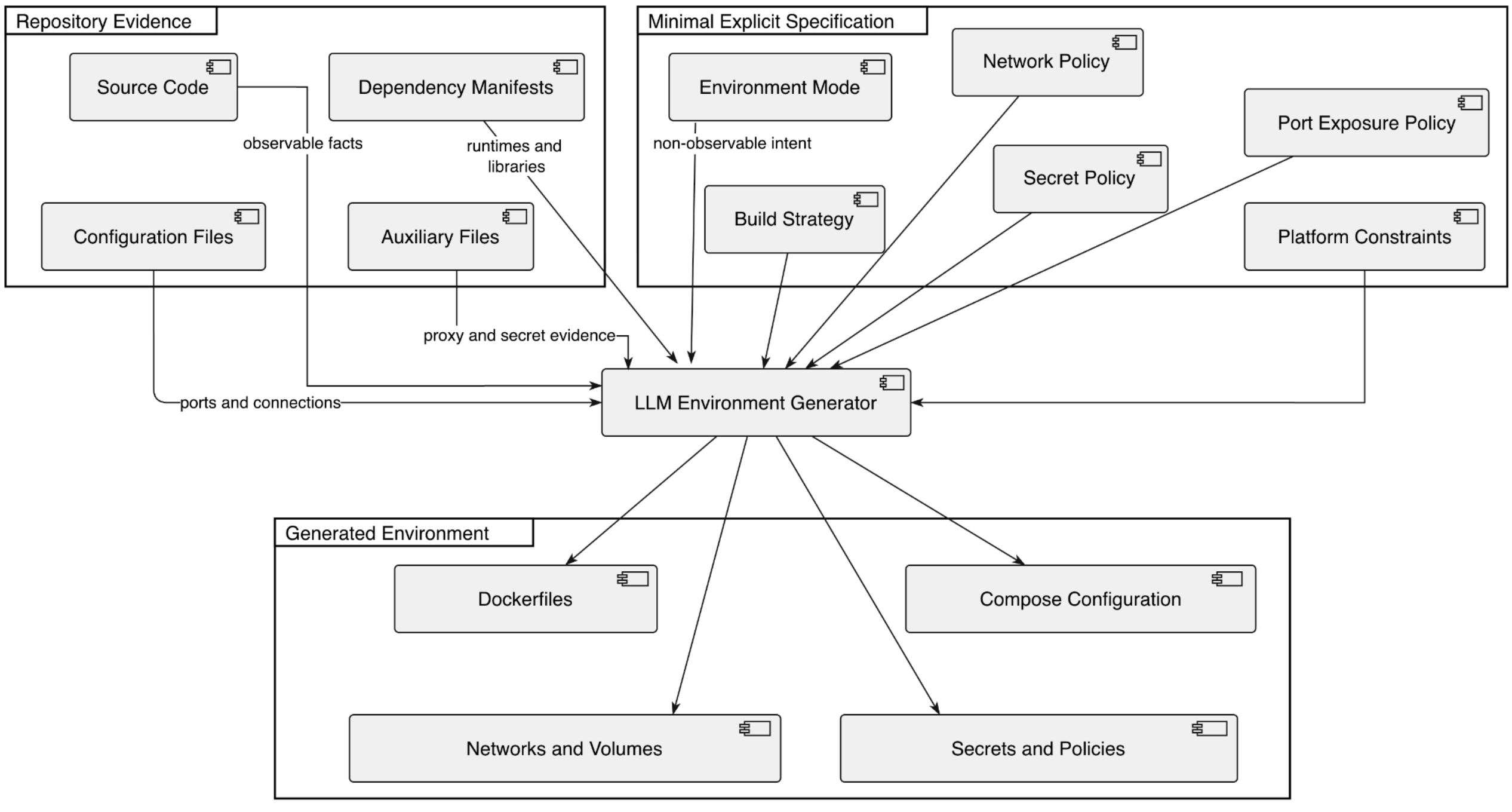

Fig. 4. From Repository-Only Generation to Specification-Driven DevOps.

As shown in Fig. 4, the proposed model supplements repository evidence with only the deployment decisions that are absent or underdetermined, rather than replacing repository-based inference with a complete manually written specification.

*Q. Section conclusion*

The experiment showed that the evaluated LLM generated executable multi-service environments from repository evidence while incompletely reconstructing workflow, security, portability, optimization, and production intent.

The generated environments achieved functional equivalence under the tested oracles without full structural or deployment-intent fidelity.

This distinction provides the empirical motivation for Specification-Driven DevOps: repository evidence should be used to infer what is observable, while a compact explicit specification should provide the deployment intent that source code cannot communicate.

The limitations affecting the interpretation and generalizability of these findings are discussed in Section VII.

## V. Repository-Level Case Studies

This section presents repository-level evidence showing how source artifacts supported generated services, dependencies, credentials, proxy relationships, and ports, and where the resulting environments diverged from developer-authored configurations.

*A. Case Study 1: `example-voting-app`*

**Repository characteristics**

The example-voting-app repository represents the most structurally heterogeneous case. It combines:

- a Python voting frontend;
- a Node.js result frontend;
- a C#/.NET background worker;
- Redis as a message queue;
- PostgreSQL as persistent storage.

The application therefore contains five operational services:

$$S_{vote} = \{s_{vote}, s_{result}, s_{worker}, s_{redis}, s_{db}\}. \quad (46)$$

The repository represents three programming-language ecosystems, an asynchronous processing chain, a background worker with no public HTTP endpoint, and database credentials embedded directly in `Program.cs`.

Table XI maps these repository signals to the inferred services and generated deployment elements.

TABLE XI. REPOSITORY EVIDENCE AND INFERRED DEPLOYMENT ELEMENTS FOR EXAMPLE-VOTING-APP

| Repository evidence | Inferred fact | Generated deployment element | Experimental result |
|---|---|---|---|
| Python voting application | Browser-facing voting service | `vote` application service and Python Dockerfile | Correct |
| Node.js result application | Browser-facing result service | `result` service and Node.js Dockerfile | Correct |
| .NET background-processing code | Non-HTTP worker service | Separate `worker` service | Correct, 1/1 |
| Redis client and queue use | Redis is required for asynchronous messaging | `redis` infrastructure service | Correct |
| PostgreSQL access in worker/result path | Relational database is required | `db`/PostgreSQL service | Correct |
| Hardcoded credentials in `Program.cs` | Database credential values | Matching database and worker configuration | Correct |
| Application listening ports | Two public interfaces | Host mappings for voting and result services | Correct |
| Inter-service references | Processing order and dependencies | Correct service hostnames and `depends_on` relationships | Correct |

The evidence mappings in Table XI establish the basis for reconstructing the complete voting-application topology.

**Inferred service topology**

The generated environment correctly reconstructed the application pipeline:

$$vote \rightarrow redis \rightarrow worker \rightarrow db \rightarrow result. \quad (47)$$

The resulting five-service topology and end-to-end vote-processing path are shown in Fig. 5.

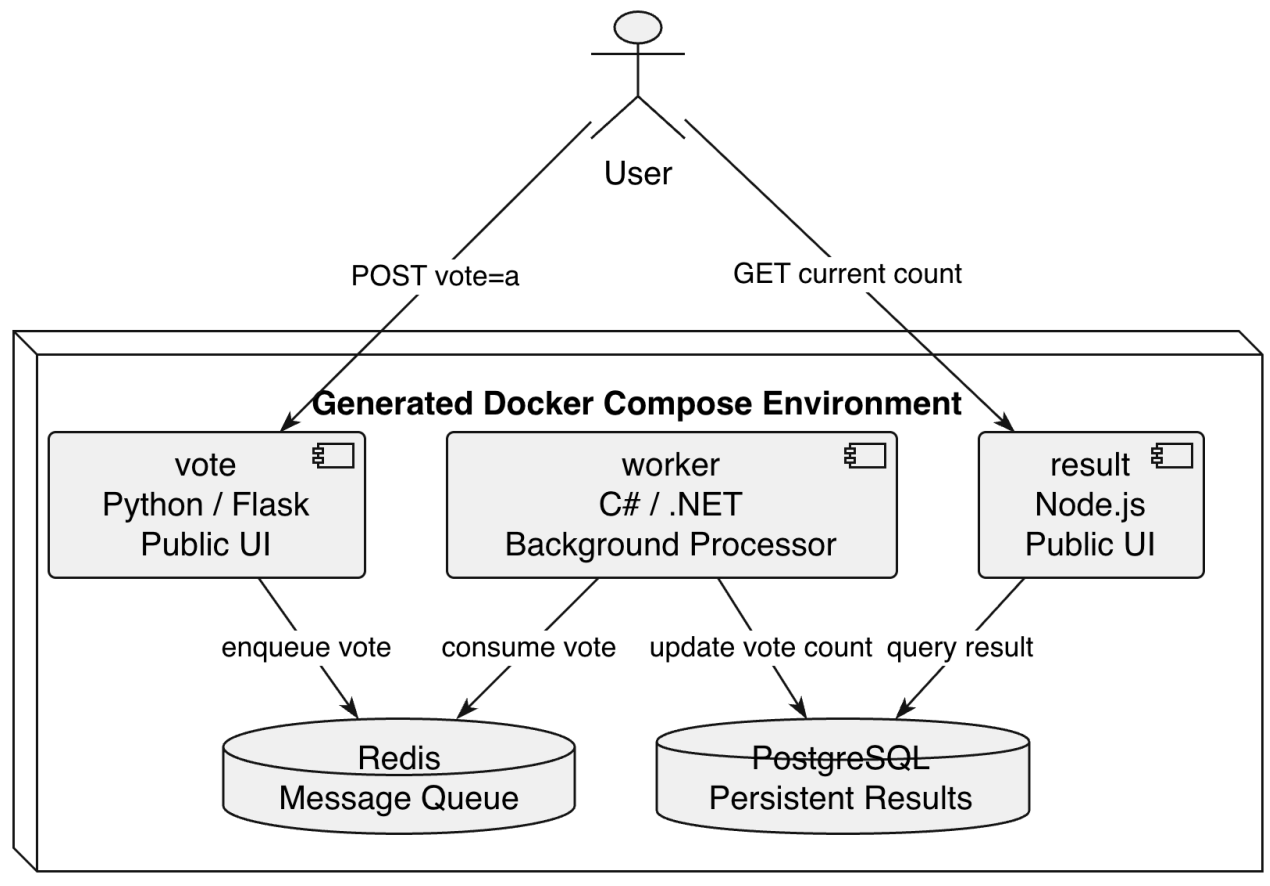

Fig. 5. Repository-Level Case: example-voting-app.

As illustrated in Fig. 5, the worker connects the Redis queue to PostgreSQL without exposing an HTTP endpoint. Worker identification therefore demonstrates that service inference was not limited to processes exposing HTTP ports.

**Credential propagation**

The database credentials were explicitly available in the source code. Therefore, the transformation followed the observable-value case:

$$q_{source} \rightarrow q_{database} \wedge q_{worker}. \quad (48)$$

The generated configuration propagated the source-visible credential values consistently to the PostgreSQL and .NET worker services.

The experimentally confirmed relationship can be represented schematically as follows; the fragment illustrates credential propagation and is not a verbatim reproduction of the generated Compose file:

```
services:
  db:
    environment:
POSTGRES_USER: <repository-derived-user>
POSTGRES_PASSWORD: <repository-derived-password>

  worker:
    environment:
POSTGRES_USER: <repository-derived-user>
POSTGRES_PASSWORD: <repository-derived-password>
      POSTGRES_HOST: db
```

The relevant result is not the credential value itself but its consistent propagation across dependent services.

**Functional validation**

The generated environment passed the voting-application oracle defined in Table IV, confirming execution of the complete Flask–Redis–.NET worker–PostgreSQL–Node.js pipeline.

**Divergences and interpretation**

The environment achieved functional equivalence while retaining the workflow, build, network, and portability differences summarized in Table VIII.

*B. Case Study 2: react-rust-postgres*

**Repository characteristics**

The second repository contains:

- a React frontend;
- a Rust backend based on Actix Web;
- a PostgreSQL database.

Its service set is:

$$S_{rust} = \{s_{frontend}, s_{backend}, s_{postgres}\}. \quad (49)$$

This repository represents three challenging inference conditions:

- database credentials were present only in the excluded original Compose file;
- the frontend proxy configuration was hidden in setupProxy.js;
- the repository contained a port ambiguity involving the frontend application port and the port used in the developer container configuration.

Table XII summarizes the repository evidence, inferred facts, generated deployment elements, and experimental outcomes for this case.

TABLE XII. REPOSITORY EVIDENCE AND INFERRED ELEMENTS FOR REACT-RUST-POSTGRES

| Repository evidence | Inferred fact | Generated deployment element | Experimental result |
|---|---|---|---|
| React source and `package.json` | Frontend service | Frontend Dockerfile and service | Correct |
| Rust/Actix source and `Cargo.toml` | Backend service and Rust runtime | Backend Dockerfile and service | Correct |
| PostgreSQL client/configuration | PostgreSQL dependency | Database service | Correct |
| `setupProxy.js` | `/api/users` must be proxied to backend | Frontend-to-backend routing | Correct, 1/1 for this repository |
| Frontend server configuration | Port 4000 | Frontend port configuration | Correct |
| Environment-variable references without values | Credentials required but unavailable | New internally consistent credentials | Functionally correct |
| Updated Actix dependency constraints | Rust 1.88 required | Base-image correction | Correct after one edit |
| Backend listening port | Internal API port | Backend container port and host mapping | Functionally valid but over-exposed |

The mappings in Table XII show that most runtime properties were reconstructed successfully, while base-image compatibility and backend exposure required separate interpretation.

**Hidden proxy discovery**

The model discovered `setupProxy.js` even though the file was not identified in the prompt or referenced from an obvious main entry point. It inferred that requests under /api/users should be routed from the frontend to the Rust backend.

The evidence mapping was:

$$setupProxy.js \rightarrow /api/users \rightarrow s_{frontend} \rightarrow s_{backend}. \quad (50)$$

This result shows that repository traversal extended beyond explicitly named entry-point files. This observation is based on one repository-specific instance and therefore should not be generalized beyond the evaluated case. The inferred frontend-proxy, backend, and database request path is illustrated in Fig. 6.

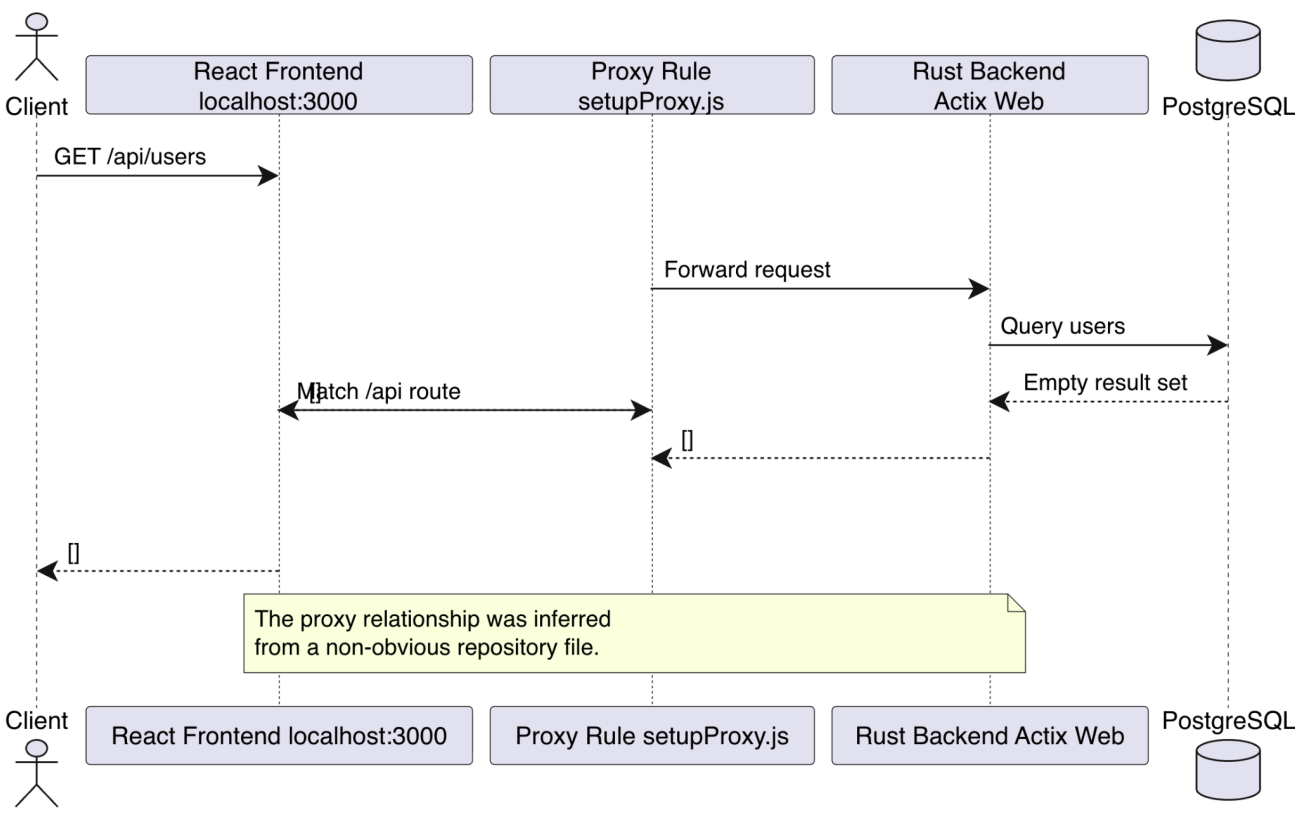

Fig. 6. Repository-Level Case: react-rust-postgres.

As shown in Fig. 6, the frontend request is routed through the discovered proxy configuration to the Rust backend and then to PostgreSQL. The case also illustrates how unavailable credential values were handled.

**Generation of underdetermined credentials**

Because the developer-authored credential values were absent from the repository input, the model generated alternative values and propagated them consistently across the PostgreSQL and Rust services. This case provides repository-level evidence for the internal-consistency analysis in Section IV.F.

**Rust version correction**

The generated Dockerfile used rust:1.85, whereas the current Actix Web dependency required Rust 1.88. Changing the base image to rust:1.88 resolved the build failure without modifying the inferred topology, dependencies, ports, credentials, or Compose relationships.

**Functional validation**

After the base-image correction, the environment passed the functional oracle defined in Table IV.

**Port over-exposure**

The developer kept the Rust backend internal, but the generated Compose configuration published its port to the host:

```
backend:
  ports:
    - "8000:8000"
```

The host mapping was unnecessary for inter-service communication and represented a less restrictive exposure policy.

### C. *Case Study 3:* `react-java-mysql`

**Repository characteristics**

The third repository consists of:

- a React frontend;
- a Java Spring Boot backend;
- a MariaDB-based developer environment, with MySQL-compatible configuration.

Its service set is:

$$S_{java} = \{s_{frontend}, s_{backend}, s_{db}\}. \quad (51)$$

The principal experimental signal was `db/password.txt`, which was used to evaluate whether the model could infer a Docker secrets mechanism from file presence and repository context alone. Table XIII summarizes the evidence-to-deployment mappings and experimental outcomes for this repository.

The evidence summarized in Table XIII shows that the model reconstructed the application services, API route, secret-file mechanism, and database configuration, while retaining deployment-intent divergences in host exposure and frontend serving.

**Docker secret inference**

The LLM encountered db/password.txt without access to the developer-authored Dockerfile or Compose definition. The model interpreted `db/password.txt` as evidence of a file-based secret mechanism and generated the corresponding Compose declaration, service mounts, and `_FILE`-based database configuration.

TABLE XIII. REPOSITORY EVIDENCE AND INFERRED ELEMENTS FOR REACT-JAVA-MYSQL

| Repository evidence | Inferred fact | Generated deployment element | Experimental result |
|---|---|---|---|
| React application | Frontend service | Frontend Dockerfile and service | Correct |
| Spring Boot application | Java backend and port 8080 | Backend Dockerfile and service | Correct |
| Database client/configuration | MySQL-compatible database required | MySQL service | Functionally correct |
| `db/password.txt` | File-based secret mechanism | Compose secret and mounted secret file | Correct, 1/1 |
| Database container conventions | `_FILE` variable must be used | `MYSQL_ROOT_PASSWORD_FILE` | Correct |
| Frontend API routing | Backend proxy target | Frontend-to-backend route | Correct |
| Java source compatibility | Java 11 sufficient | Java 11 base image | Functional but differs from Java 17 reference |
| Backend listening port | Internal API available | Port 8080 published to host | Over-exposed |
| React development command | Runnable frontend | Development server as final process | Functional but not production-ready |

The transformation was:

$$db/password.txt \to qdb \to secrets \to$$

$$/run/secrets/db - password.$$

The confirmed secret-handling relationship can be represented schematically as follows, while the complete inference and propagation path is illustrated in Fig. 7. The fragment is not a verbatim reproduction of the complete generated configuration:

```
secrets:
  db-password:
    file: ./db/password.txt

services:
  db:
    image: mysql:8.0
    environment:
                          MYSQL_ROOT_PASSWORD_FILE:
/run/secrets/db-password
    secrets:
      - db-password

  backend:
    secrets:
      - db-password
```

As illustrated in Fig. 7, the password file was interpreted as a secret declaration, mounted into the database and backend services, and consumed through the file-based environment-variable mechanism.

**Image and version alternatives**

The generated environment used MySQL 8.0 instead of the developer-authored MariaDB configuration and Java 11 instead of Java 17. Both alternatives passed the functional oracle; however, the experiment did not separately evaluate portability differences between the database images.

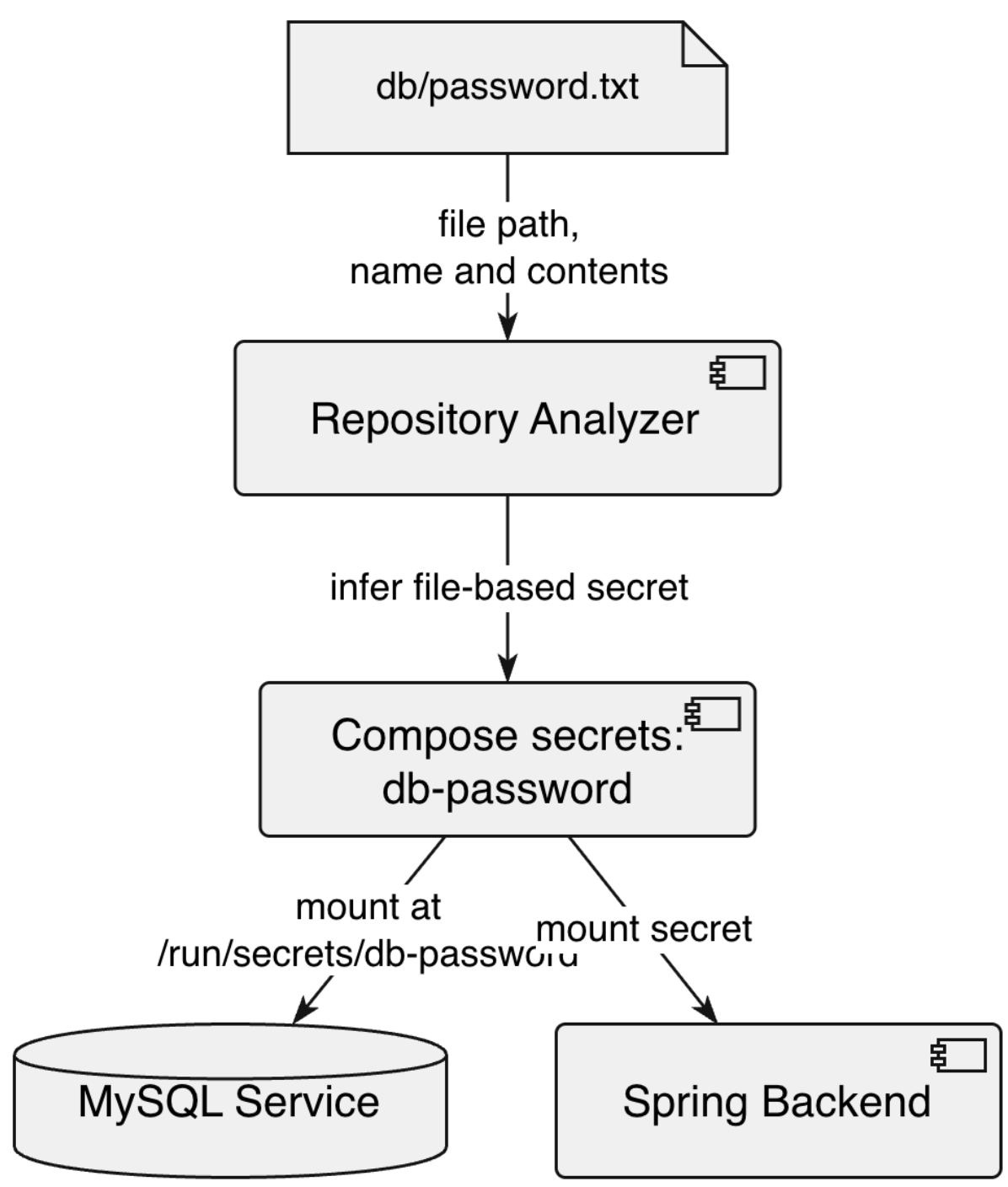

Fig. 7. Repository-Level Case: Docker Secret Inference.

**Functional validation**

The environment passed startup and the Java-application oracle without manual correction, confirming the React–Spring Boot–database execution path.

**Port over-exposure and production serving**

The generated configuration published backend port 8080:

```
backend:
  ports:
    - "8080:8080"
```

The developer intended the backend to remain internal and to be accessed through the frontend.

The generated frontend also used a development server as the final runtime process rather than a production build served through Nginx. Therefore, the environment was operational but not production-oriented.

These differences represent security-exposure and production-serving intent gaps.

*D. Cross-case synthesis*

Across the evaluated repositories, runtime-essential services and dependencies were reconstructed more reliably than policy-oriented deployment decisions. The voting application demonstrated heterogeneous service and worker inference, the Rust application demonstrated hidden proxy discovery and internally consistent handling of missing values, and the Java application demonstrated file-based secret inference.

For the evaluated repositories, repository evidence supported completion of the tested runtime pipelines but not complete reconstruction of deployment policy.

The case studies motivate the minimal explicit deployment specification developed in Section VI.

## VI. Minimal Deployment Specification for Specification-Driven DevOps

### A. Purpose of the specification

The case studies show that repository contents can provide a substantial part of the deployment model. Repeating all observable details in an external specification would create duplication and increase maintenance cost.

The specification should therefore include only the information that:

- is absent from the repository;
- is underdetermined by the repository;
- represents workflow, security, platform, or production intent;
- must not be replaced by an arbitrary model default.

These criteria define the scope of the minimal specification and prevent duplication of facts already available in repository artifacts. Following the definition introduced in Section I, the minimal specification contains only deployment knowledge that is required but cannot be recovered reliably from repository artifacts.

### B. Specification categories derived from the experiment

The specification categories were derived from experimentally observed omissions and underdetermined choices in deployment context, workflow, network policy, port exposure, build strategy, secret handling, production serving, and infrastructure constraints.

### C. Proposed deployment specification

The following YAML is a conceptual specification template derived from the observed information gaps. Its concrete values are illustrative and were not evaluated in the conducted experiment.

```
# deployment-spec.yml

deployment:
  environment: <dev|staging|production>
  platform: <target-platform>

workflow:
  live_reload: true
  debug_ports: true

network:
  segmentation: true
  public_services:
    - frontend
  internal_services:
    - backend
    - db

build:
  multi_stage: true
  named_targets:
    - dev
    - final
  cross_platform: true

secrets:
   mechanism: env_file            # env_file |
docker_secrets | hardcoded | vault
  generate:
                                           <secret-name>:
<generation-command-or-provider>

production:
   serving: nginx             # nginx | caddy |
none
  static_build_command: "npm run build"

infrastructure:
  versions: {...}
  persistence: {...}
  resource_limits: {...}
```

The template omits services, ports, dependencies, and proxy routes that were successfully inferred in the evaluated repositories.

### D. Formal specification model

The minimal specification can be represented as:

$$S_{min} = \langle D, W, N, P, B, Q, R, I\rangle, \quad (52)$$

where $D$ is deployment context; $W$ is workflow configuration; $N$ is network topology; $P$ is port-exposure policy; $B$ is build strategy; $Q$ is secret-management policy; $R$ is production-serving strategy; $I$ is infrastructure constraints.

**Deployment context**

Deployment context: $D = \langle e, a\rangle$,

where

$$e \in \{dev, staging, production\},$$

and $a$ is the target architecture, such as:

$$a \in \{linux/amd64, linux/arm64\}.$$

This category controls version and image selection as well as development versus production behavior.

Network policy:

$$N = \langle g, S_{public}, S_{internal}\rangle, \quad (53)$$

where $g \in \{0, 1\}$ indicates whether segmentation is required; $S_{public}$ is the set of host-accessible services; $S_{internal}$ is the set of services that must remain internal.

The sets must satisfy:

$$S_{public} \cap S_{internal} = \emptyset. \quad (54)$$

Build strategy: $B = \langle m, T, c\rangle$,

where $m$ indicates whether multi-stage builds are required; $T$ is the set of named build targets; $c$ indicates whether cross-platform build arguments must be included.

Secret policy $Q = \langle q_m, Q_g\rangle$, where $q_m$ is selected from:

$$q_m \in \{env_file, docker_secrets, hardcoded, vault\}, \quad (55)$$

and $Q_g$ is the set of values that must be generated during deployment.

### E. Mapping Observed Omissions to Proposed Specification Fields

As summarized in Table XIV, the specification fields are intended to constrain network segmentation, host exposure, workflow mode, build structure, production serving, platform selection, credential policy, and version selection.

Table XIV maps each experimentally observed omission or underdetermined choice to a proposed specification field and intended generation constraint.

TABLE XIV. MAPPING OBSERVED OMISSIONS TO PROPOSED SPECIFICATION FIELDS

| Experimental result | Specification field | Intended generation constraint |
|---|---|---|
| Flat default network in 3/3 repositories | network.segmentation | Request explicit named-network generation |
| Backend ports exposed in 2/2 applicable cases | public_services, internal_services | Constrain host publication to declared public services |
| No live-reload mounts in 3/3 cases | workflow.live_reload | Request development source mounts and commands when live reload is enabled |
| No multi-stage builds in 3/3 cases | build.multi_stage | Request separate build and runtime stages |
| No production Nginx in 2/2 cases | production.serving | Declare the required production-serving mechanism |
| No cross-platform arguments in 3/3 cases | build.cross_platform, deployment.platform | Request platform-aware base images and build arguments |
| Credentials invented when unavailable | secrets.mechanism, secrets.generate | Declare an explicit credential-generation or secret-store policy |
| Java/database version differences | infrastructure.*_version | Declare required runtime and infrastructure versions |

The following process, conflict-resolution principles, validation constraints, and traceability extensions constitute a proposed framework and were not executed in the conducted experiment.

### F. Proposed Specification-Driven Generation Process

The proposed generation process combines repository-derived facts with explicitly declared intent, as shown in Fig. 8. In the proposed process, each deployment decision is supported by repository evidence, explicit specification, or is marked as an ungrounded default for review.

### G. Proposed Conflict-Resolution Principles

Repository evidence and explicit specification may conflict. The following conflict-resolution principles are proposed.

**Principle 1: explicit policy should override an inferred default**

If the repository shows that a backend listens on port 8080, but the specification declares it internal:

$$Listen(8080) + InternalOnly \Rightarrow ExposeContainer(8080) \wedge \neg PublishHost(8080).$$

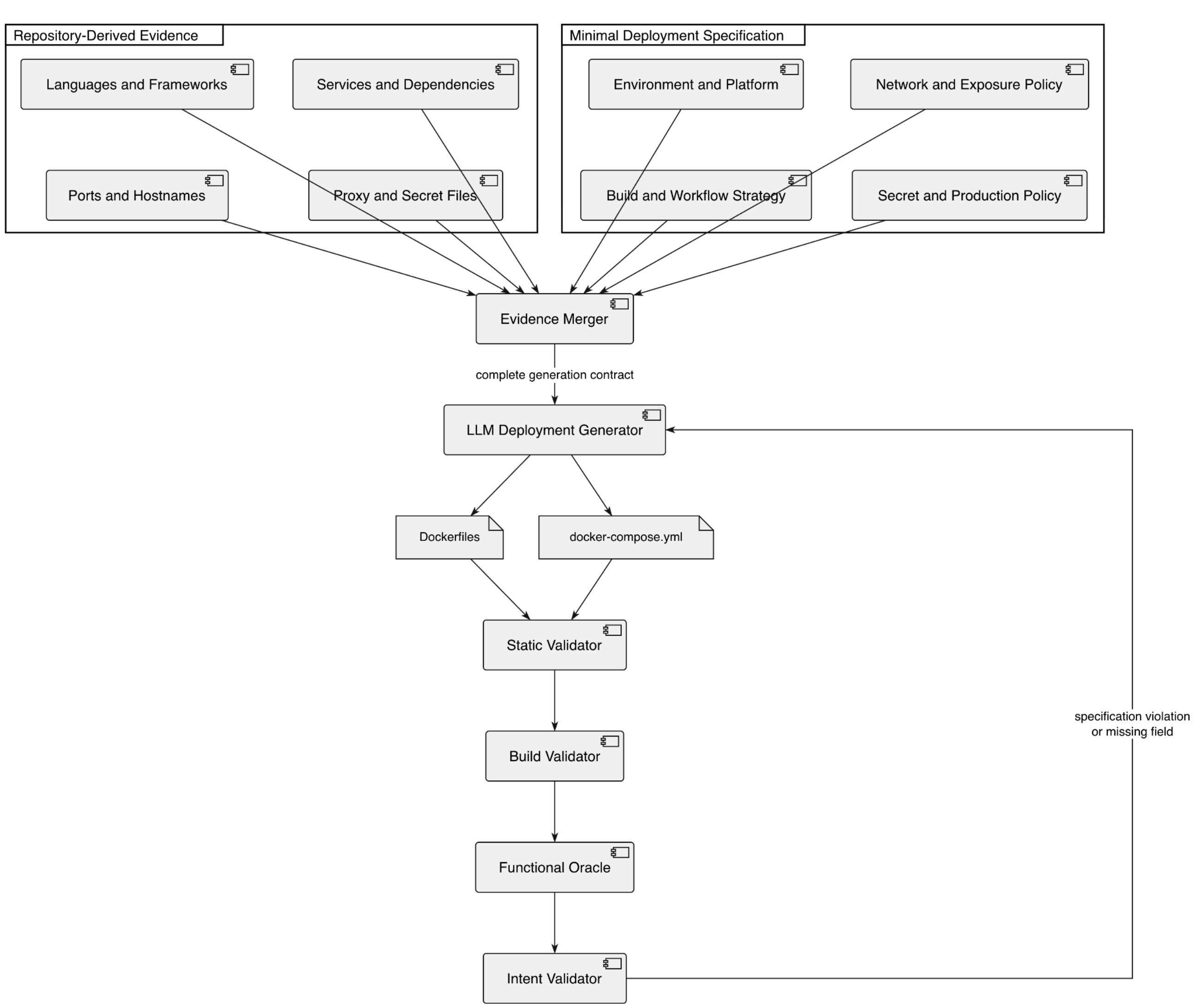

Fig. 8. Specification-driven generation and validation cycle.

**Principle 2: a specified runtime version should override a compatible inferred alternative**

If the repository is compatible with Java 11 but the specification requires Java 17:

$$Compatible(11) \land Specified(17) \Rightarrow Use(17).$$

**Principle 3: explicit secret policy should override repository-visible secret handling**

If a password is hardcoded in source but the specification requires Docker secrets:

$$q_{hardcoded} + q_m = docker_secrets \Rightarrow Externalize(q).$$

**Principle 4: functional success should not be treated as security-policy compliance**

A passing functional oracle does not justify publishing an internal port:

$$OraclePass \neq PolicyPass.$$

As illustrated in Fig. 8, the proposed process merges repository evidence and the minimal specification before artifact generation and subsequently applies static, build, functional, and intent-oriented validation.

*H. Proposed Specification Validation Rules*

Before generation, the minimal specification should be checked for internal consistency. Table XV defines the core validation constraints for public/internal classification, production serving, live reload, debugging, secret mechanisms, build targets, platform compatibility, and service references.

Formally:

$$Valid(S_{min}) = \Lambda c_k, \tag{56}$$

where $c_k$ represents one validation constraint.

Experimental validation of the specification requires a controlled comparison between repository-only generation $G(R)$ and repository-plus-specification generation $G(R, S_{min})$. This comparison is outside the scope of the present study.

TABLE XV. CORE VALIDATION CONSTRAINTS

| Constraint | Validation rule |
|---|---|
| Public/internal overlap | A service cannot be in both lists |
| Production serving | If `environment=production`, development server must not be the default unless explicitly allowed |
| Live reload | Must be disabled for production unless explicitly permitted |
| Debug ports | Must be disabled for production unless explicitly permitted |
| Secret mechanism | Required generated values must match the selected mechanism |
| Named build target | Every target referenced by Compose must exist in the Dockerfile |
| Platform | Selected images must support the requested architecture |
| Resource limits | Referenced service names must exist in the inferred service set |

The constraints in Table XV prevent contradictory or incomplete specification values from being propagated into the generated deployment artifacts.

*I. Traceability model*

Every generated environment element should be connected to its evidence source.

For each generated deployment attribute $a_i$, the traceability function records its origin type and supporting repository artifact or specification field:

$$\tau(a_i) = \langle o_i, r_i \rangle, \tag{57}$$

where $o_i$ is the origin type; $r_i$ is the supporting repository artifact or specification field.

The origin type is:

$$o_i \in \{repository, specification, default\}. \tag{58}$$

Repository-origin examples are grounded in the experiment, whereas specification-origin examples illustrate the proposed model and were not generated experimentally. Representative traceability records are presented in Table XVI.

TABLE XVI. ILLUSTRATIVE TRACEABILITY RECORDS FOR THE PROPOSED MODEL

| Generated element | Origin | Supporting evidence |
|---|---|---|
| Redis service | Repository | Redis client usage and connection hostname |
| Worker service | Repository | .NET background-processing code |
| Frontend public port | Repository + specification | Repository-visible listening port and public services |
| Backend internal-only policy | Specification | `network.internal_services` |
| Docker secret | Repository | `db/password.txt` |
| Nginx final stage | Specification | `production.serving: nginx` |
| Rust 1.88 image | Repository + validation | Cargo dependency minimum version |
| Platform-compatible database image | Specification | `deployment.platform: linux/arm64` |

As shown in Table XVI, each generated element can be associated with repository evidence, a specification field, or a validation result. Elements classified as defaults should be included in a generation report because they represent decisions not grounded in explicit evidence.

*J. Discussion*

The minimal specification formalizes a hybrid principle: infer repository-supported facts and explicitly specify irreducible deployment intent. The model avoids two extremes.

The first extreme is repository-only generation: $G(R)$, which is efficient but may silently introduce unsafe or unsuitable defaults.

The second extreme is a complete hand-written deployment specification:

$$G(S_{complete}), \tag{59}$$

which duplicates information already encoded in source code.

Specification-Driven DevOps uses:

$$G(R, S_{min}), \tag{60}$$

where $S_{min}$ contains only irreducible intent.

The repository-level case studies showed successful runtime-topology reconstruction and incomplete deployment-policy recovery across the three evaluated applications.

The proposed specification represents deployment intent that was absent or underdetermined in repository evidence. Because specification-augmented regeneration was not conducted, the study establishes its analytical structure but not its effectiveness.

Future work should evaluate $G(R)$ and $G(R, S_{min})$ using the same repositories, functional oracles, and structural-intent criteria.

## VII. Limitations of the Conducted Experiment

The interpretation of the results is constrained by the sample size, repository selection, model configuration, interaction setting, validation scope, and number of observed cases.

### *A. Limited sample size*

The experiment evaluated three repositories:

- example-voting-app;
- react-rust-postgres;
- react-java-mysql.

All experiments used GPT Codex 5.3 through GitHub Copilot. The available interaction record does not fully document every detail required for exact replication.

The result of three functionally operational environments out of three evaluated cases applies only to this sample. It should not be interpreted as an expected success rate for arbitrary multi-service repositories. The repositories also contain relatively simple service topologies. Each application includes two or three application services together with supporting infrastructure such as Redis, PostgreSQL, MySQL, or MariaDB.

The experiment does not cover large microservice ecosystems, service meshes, complex event-driven systems, large monorepositories, or distributed applications with dozens of independently configured services. The results apply only to the three evaluated repositories and should not be generalized as evidence of universal reliability.

### *B. Repository-selection and potential-familiarity bias*

All three applications were obtained from Docker-maintained repositories. example-voting-app is provided through Docker Samples, while the other two applications originate from the Docker awesome-compose collection. These repositories are intended to demonstrate containerized application patterns and therefore have relatively clear structures and discoverable relationships among services.

This creates an application-selection bias. The evaluated repositories contain:

- clearly separated frontend, backend, worker, and database components;
- conventional dependency manifests;
- recognizable framework structures;
- discoverable ports and connection settings;
- comparatively limited configuration complexity.

Real enterprise repositories may contain:

- obsolete or conflicting configuration files;
- undocumented environment variables;
- shared libraries across services;
- conditional service startup;
- multiple runtime profiles;
- generated code;
- private dependencies;
- legacy scripts;
- incomplete or misleading documentation.

The selected repositories may overestimate performance relative to less structured enterprise repositories.

The evaluated applications are publicly available and widely used as Docker examples. Therefore, it cannot be excluded that the model had encountered similar repository structures or deployment configurations during training.

The experiment removed existing Dockerfiles and Compose files from the input, but the source material does not establish whether the model could indirectly rely on memorized patterns associated with these public repositories. Potential familiarity with the public repositories could not be excluded or quantified.

Therefore, the results should be interpreted as repository-conditioned generation outcomes, without making a strong claim that all generated elements were derived exclusively from unseen source evidence.

### *C. Model and interaction-setting limitations*

The experiment used GPT Codex 5.3 via GitHub Copilot, but the available material does not fully document every interaction detail required for exact replication.

Different models may vary in their ability to:

- traverse repository structures;
- identify service boundaries;
- infer proxy configurations;
- recognize secret-management patterns;
- select compatible base-image versions;
- preserve consistency across generated files;
- distinguish public from internal services.

The correct discovery of setupProxy.js and db/password.txt should therefore be interpreted as observed behavior of the evaluated model, not as a guaranteed capability of LLMs in general.

The following aspects are not fully specified in the experimental report:

- the complete generation prompt;
- all intermediate model messages;
- whether the model had access to terminal execution during generation;
- whether it inspected build errors before finalizing the files;
- the complete sequence of repository-navigation actions;
- the level of internet or external repository access.

The incomplete interaction record limits exact reproducibility.

*D. Limited repetitions and low-count observations*

The experiment reports one generation attempt for each repository and therefore does not measure output stability or reproducibility across repeated runs.

LLM-generated deployment artifacts may vary in:

- service names;
- image versions;
- credential values;
- build-stage structure;
- port mappings;
- network definitions;
- dependency ordering;
- runtime commands.

Because no repeated runs were performed, the stability of these output characteristics could not be assessed. For example, it remains unknown whether repeated generations would consistently:

- infer the Docker secrets mechanism;
- discover the hidden proxy configuration;
- identify the background worker;
- over-expose backend ports;
- omit network segmentation;
- select the same image versions.

Several notable observations are supported by only one or two applicable cases.

The experiment reported:

- background worker identification: 1/1;
- Docker secrets inference: 1/1;
- hidden proxy inference: 2/2;
- production-serving omission: 0/2;
- restrictive backend port exposure: 0/2.

These low-count observations should be treated as case-specific rather than statistically stable.

The observed secret-file and proxy inferences demonstrate capability in the evaluated cases but do not establish reliability across broader repository populations.

*E. Validation-scope limitations*

Each repository was validated using one deterministic end-to-end HTTP request. This is stronger than checking only YAML syntax, image build success, or container startup because it verifies the principal multi-service execution path. However, the functional oracle remains limited in scope.

The conducted oracles verified:

- service startup;
- basic service discovery;
- inter-service connectivity;
- database or message-broker access;
- correct processing of one successful request;
- expected response production.

They did not verify:

- invalid-input handling;
- authentication;
- authorization;
- error propagation;
- retry behavior;
- service recovery;
- persistence after restart;
- concurrent requests;
- performance under load;
- fault tolerance;
- security hardening;
- long-running stability.

Oracle success demonstrates execution of the tested path, not complete correctness or production readiness. The empty array returned by react-rust-postgres, for example, confirmed connectivity and successful backend execution, but did not test data modification, failure handling, or transaction behavior. The experiment identified a security-relevant tendency: backend services were published to the host in both applicable frontend–backend applications. However, no dedicated security tests were conducted.

The study did not evaluate:

- unauthorized network access;
- credential leakage;
- container privileges;
- vulnerable images;
- secret exposure in logs;
- open database ports;
- default credentials;
- TLS or mTLS;
- image provenance;
- supply-chain risks.

Consequently, the identified security issues represent structural divergences rather than results of penetration testing or vulnerability assessment. The study identifies structural security-related divergences but does not quantify security. The generated environments were tested for basic functionality only. No measurements were reported for:

- startup time;
- image size;
- build duration;
- cache reuse;
- CPU consumption;
- memory consumption;
- throughput;
- response latency.

Consequently, the observed omissions of layer caching and multi-stage builds were identified structurally rather than through measured performance degradation; resource limits were not evaluated.

The experiment identified absent optimizations but did not measure their performance or resource impact. Taken together, these limitations characterize the study as an exploratory feasibility experiment. The findings apply to the three evaluated repositories and do not establish general reliability, production readiness, generation stability, or the effectiveness of specification-augmented regeneration.

*F. One build-level manual correction*

One environment required a manual base-image change from Rust 1.85 to 1.88. The experiment therefore achieved a first-build success rate of 2/3 and a final operational success rate of 3/3, not three autonomous first-attempt successes.

*G. Manual structural comparison*

Generated and developer-authored artifacts were compared manually across image versions, database technologies, networks, build strategies, volumes, production serving, and host-port exposure. This introduced evaluator judgment. Some differences were classified as harmless, while others were interpreted as security-, workflow-, or production-related deviations.

These interpretations are reasonable, but the experiment did not use a formally predefined scoring rubric or multiple independent evaluators. Structural fidelity should therefore be treated as a qualitative result rather than a fully objective numerical measure.

### H. Credential inference was tested under favorable conditions

The three repositories represent three credential patterns:

- hardcoded credentials;
- credentials absent from the repository input;
- credentials stored in a dedicated file.

These cases are useful, but the experiment does not cover more complex credential scenarios. It does not evaluate repositories where:

- only DATABASE_URL is referenced without a default;
- credentials are injected exclusively through CI/CD;
- vault references are used;
- secrets are encrypted;
- cloud identity replaces passwords;
- multiple environments use different credential schemes.

The credential findings apply only to the three evaluated patterns.

### I. No iterative regeneration was evaluated

The study did not evaluate an automated iterative correction loop in which the LLM receives build or runtime errors and regenerates the deployment artifacts. The Rust version mismatch was corrected manually rather than through a documented LLM-driven feedback cycle. Consequently, the experiment does not provide results concerning:

- automated error interpretation;
- multi-step regeneration;
- convergence speed;
- number of required iterations;
- whether iterative feedback improves correctness;
- whether iteration introduces new configuration errors.

The experiment evaluated initial generation with limited manual correction, not autonomous iterative repair.

### J. The proposed specification was derived but not experimentally validated

The study derives a minimal deployment specification from the recurring omissions observed in the three experiments. The proposed fields address:

- deployment environment;
- network segmentation;
- public and internal services;
- port exposure;
- build strategy;
- secret mechanism;
- production serving;
- platform and infrastructure constraints.

However, the experiment did not include a second generation phase using this specification.

The experiment supports derivation of the specification categories but not claims that the specification improved generation. No experimental evidence was produced for improvements in network segmentation, port exposure, multi-stage generation, Nginx serving, or platform compatibility after applying the proposed specification. The proposed `deployment-spec.yml` is an analytically derived template rather than a validated generation mechanism.

## VIII. Conclusions

The repository-to-environment transformation was formalized as a mapping from source code, dependency manifests, configuration files, credential-related artifacts, auxiliary files, and repository structure to application and infrastructure services, Dockerfiles, Compose configurations, networks, volumes, secrets, and port policies. The transformation produced runnable environments for all three evaluated repositories.

The experiments identified a practical boundary between repository-inferable runtime knowledge and deployment intent that was absent or underdetermined in repository evidence. The model correctly identified service topology, application ports, infrastructure dependencies, service hostnames, dependency ordering, a background worker, hidden proxy configurations, and a file-based Docker secrets pattern. In contrast, network segmentation, multi-stage builds, layer-caching strategy, live-reload mounts, production frontend serving, restrictive host-port exposure, and cross-platform build logic were not reconstructed.

The DevOps intent gap was defined as the difference between deployment knowledge required for the intended environment and knowledge observable or derivable from repository artifacts. The conducted experiments demonstrated that repository contents provide a partial source of deployment knowledge rather than a complete representation of deployment intent. The generated environments supported execution but omitted trust-boundary, exposure, environment-mode, production-serving, build, and platform decisions.

Deterministic end-to-end HTTP oracles validated the complete multi-service execution path for each repository. All three applications passed after one version-level correction in the Rust case.

A minimal explicit deployment specification was derived from recurring information gaps that could not be resolved reliably from repository artifacts. The resulting specification covers deployment context and workflow, network, exposure, build, secret, production, and platform policies that were absent or underdetermined in repository evidence. The specification was analytically derived from the observed gaps and was not validated through regeneration.

The evaluation distinguished functional correctness from structural and deployment-intent fidelity. All three environments became operational, although one required a Rust base-image correction; therefore, the first-build and final operational success rates were 2/3 and 3/3, respectively. At the same time, repeated structural omissions showed that functional equivalence with the developer environment does not imply complete structural, security, workflow, or production fidelity.

Manual comparison showed that internally consistent alternative credentials and compatible runtime versions could preserve functionality, whereas flat networking, unnecessary backend host exposure, absent multi-stage builds, and missing production frontend serving represented deployment-intent divergences. For the evaluated applications, repository artifacts supported reconstruction of runtime requirements, whereas production-oriented generation additionally requires explicit representation of deployment decisions that are absent or underdetermined in source artifacts.